\documentclass[12pt]{article}
\pdfoutput=1

\usepackage{amsmath,amssymb,amsfonts,epsfig,cite,setspace,bigstrut,framed,eufrak}
\usepackage[all]{xy}
\usepackage{color}
\usepackage{pifont}

\makeatletter \@addtoreset{equation}{section} \makeatother
\renewcommand{\theequation}{\thesection.\arabic{equation}}

\begin{document}

\vskip 0.25in

\newcommand{\todo}[1]{{\bf\color{blue} !! #1 !!}\marginpar{\color{blue}$\Longleftarrow$}}
\newcommand{\nn}{\nonumber}
\newcommand{\comment}[1]{}
\newcommand\T{\rule{0pt}{2.6ex}}
\newcommand\B{\rule[-1.2ex]{0pt}{0pt}}

\newcommand{\CO}{{\cal O}}
\newcommand{\cI}{{\cal I}}
\newcommand{\cM}{{\cal M}}
\newcommand{\cW}{{\cal W}}
\newcommand{\cN}{{\cal N}}
\newcommand{\cR}{{\cal R}}
\newcommand{\cH}{{\cal H}}
\newcommand{\cK}{{\cal K}}
\newcommand{\cT}{{\cal T}}
\newcommand{\cZ}{{\cal Z}}
\newcommand{\cO}{{\cal O}}
\newcommand{\cQ}{{\cal Q}}
\newcommand{\cB}{{\cal B}}
\newcommand{\cC}{{\cal C}}
\newcommand{\cD}{{\cal D}}
\newcommand{\cE}{{\cal E}}
\newcommand{\cF}{{\cal F}}
\newcommand{\cA}{{\cal A}}
\newcommand{\cX}{{\cal X}}
\newcommand{\IA}{\mathbb{A}}
\newcommand{\IP}{\mathbb{P}}
\newcommand{\IQ}{\mathbb{Q}}
\newcommand{\IH}{\mathbb{H}}
\newcommand{\IR}{\mathbb{R}}
\newcommand{\IC}{\mathbb{C}}
\newcommand{\IF}{\mathbb{F}}
\newcommand{\IV}{\mathbb{V}}
\newcommand{\II}{\mathbb{I}}
\newcommand{\IZ}{\mathbb{Z}}
\newcommand{\re}{{\rm Re}}
\newcommand{\im}{{\rm Im}}
\newcommand{\tr}{\mathop{\rm Tr}}
\newcommand{\ch}{{\rm ch}}
\newcommand{\rk}{{\rm rk}}
\newcommand{\ext}{{\rm Ext}}
\newcommand{\bi}{\begin{itemize}}
\newcommand{\ei}{\end{itemize}}
\newcommand{\beq}{\begin{equation}}
\newcommand{\eeq}{\end{equation}}
\newcommand{\bea}{\begin{eqnarray}}
\newcommand{\eea}{\end{eqnarray}}
\newcommand{\ba}{\begin{array}}
\newcommand{\ea}{\end{array}}

\newcommand{\CN}{{\cal N}}
\newcommand{\y}{{\mathbf y}}
\newcommand{\z}{{\mathbf z}}
\newcommand{\C}{\mathbb C}\newcommand{\R}{\mathbb R}
\newcommand{\CA}{\mathbb A}
\newcommand{\CP}{\mathbb P}
\newcommand{\cP}{\mathcal P}
\newcommand{\tmat}[1]{{\tiny \left(\begin{matrix} #1 \end{matrix}\right)}}
\newcommand{\mat}[1]{\left(\begin{matrix} #1 \end{matrix}\right)}
\newcommand{\diff}[2]{\frac{\partial #1}{\partial #2}}
\newcommand{\gen}[1]{\langle #1 \rangle}

\newtheorem{theorem}{\bf THEOREM}
\newtheorem{proposition}{\bf PROPOSITION}
\newtheorem{observation}{\bf OBSERVATION}

\def\theequation{\thesection.\arabic{equation}}
\newcommand{\setall}{
	\setcounter{equation}{0}
}
\renewcommand{\thefootnote}{\fnsymbol{footnote}}

\begin{titlepage}
\vfill
\begin{flushright}
{\tt\normalsize KIAS-P16012}\\

\end{flushright}
\vfill

\begin{center}
{\Large\bf Witten Index for Noncompact Dynamics}

\vskip 1.5cm

Seung-Joo Lee\footnote{\tt seungsm@vt.edu} and
Piljin Yi\footnote{\tt piljin@kias.re.kr}
\vskip 5mm
{\it $^*$Department of Physics, Robeson Hall, Virginia Tech, \\
Blacksburg, VA 24061, U.S.A.
}
\vskip 3mm
{\it $^\dagger$School of Physics and Quantum Universe Center \\
Korea Institute for Advanced Study, Seoul 02455, Korea}

\end{center}
\vfill

\begin{abstract}
Among gauged dynamics motivated by string theory,
we find many with gapless asymptotic directions. Although the natural
boundary condition for ground states is $L^2$, one often turns on
chemical potentials or supersymmetric mass terms to regulate the
infrared issues, instead, and computes the twisted partition function.
We point out how this procedure generically fails to capture physical
$L^2$ Witten index with often misleading results. We also explore how,
nevertheless, the Witten index is sometimes intricately embedded in
such twisted partition functions. For $d=1$ theories with gapless
continuum sector from gauge multiplets, such as non-primitive quivers
and pure Yang-Mills, a further subtlety exists, leading to fractional
expressions. Quite unexpectedly, however, the integral $L^2$ Witten
index can be extracted directly and easily from the twisted partition
function of such theories. This phenomenon is tied to the notion
of the rational invariant that appears naturally in the wall-crossing
formulae, and offers a general mechanism of reading off Witten index
directly from the twisted partition function. Along the way,
we correct early numerical results for some of $\CN=4,8,16$ pure
Yang-Mills quantum mechanics, and count threshold bound states for
general gauge groups beyond $SU(N)$.
\end{abstract}

\vfill
\end{titlepage}

\tableofcontents
\renewcommand{\thefootnote}{\#\arabic{footnote}}
\setcounter{footnote}{0}
\vskip 2cm

\section{Index vs. Twisted Partition Function}

The notion of index of elliptic operators has served theoretical
physics in countless problems in supersymmetric context. The relevant
object in the latter language is the Witten index~\cite{Witten:1982df},
defined as a trace over the physical Hilbert space,
$$
\cI=\lim_{\beta\rightarrow \infty}{\rm tr}
\left[ (-1)^{F}\cdots \, e^{-\beta H}\right] \ ,
$$
where $\beta$ is the Euclidean time span, $H=Q^2$ for some choice
of supercharge $Q$, and the ellipsis denotes any operators that
commute with $Q$, $H$, and $(-1)^{F}$.
Its key property, guaranteed by integral nature of the quantity,
is the invariance under ``small"  deformations.
Since ``small" deformation can only induce a continuous change,
an integral quantity like this cannot be affected. Hence,
one is free to modify the elliptic operator in question, or
alternatively the supersymmetric dynamics in question, in any
way as long as the underlying chirality or supersymmetric
property is intact. For example, if the dynamics has a gap,
one can choose to scale it up to an arbitrarily large value
and make the spectrum completely discrete, without affecting
the value of the index.

In a more modern form, this type of thinking leads to the
so-called localization procedure, where one performs the
deformation at the level of path-integral, without having to
go to the Hamiltonian formulation. The localization works
because of two key facts. The first is an introduction
of ``chemical potentials" as  complex parameters that carry
the grading information under global symmetries \cite{AlvarezGaume:1986nm},
and the second is the identification of a convenient supercharge
which acts much like the old-fashioned BRST operators \cite{Witten:1982im,Moore:1998et}.
The former
can introduce various gaps to the index problem and should not
be thought of as a mere reservoir of charge information.
The latter then allows deformations of the dynamics that
scale up the gaps. While these two elements are hardly new
as mathematical ideas, systematic institution at the level of
path-integral was quite recent, leading to many new computational
tools and results.

However, such conveniences can often be taken too far if one
fails to appreciate what is meant by ``small'' deformation.
A well-known example of this subtlety can be found in
supersymmetric quantum mechanics with four supercharges
or less, where wall-crossing phenomena occur. For quantum mechanics
associated with $d=4$ BPS dyons, such wall-crossing
phenomenon~\cite{Bak:1999da,Bak:1999ip,Gauntlett:1999vc,Stern:2000ie,Denef}
reflects that of $d=4$ $\CN=2,4$ theories~\cite{Seiberg:1994rs,Seiberg:1994aj,Ferrari:1996sv,Bergman:1997yw,Lee:1998nv}.
For a gauged quantum mechanics \cite{Denef}, its Fayet-Iliopoulos (FI) constant $\zeta$,
if present in the theory, can change the Witten index discontinuously,
as $\zeta$ is dialed continuously across the $\zeta=0$ wall.
While a naive invariance argument seemingly applies
to the continuous parameter $\zeta$,  one can see upon
closer inspection that setting $\zeta=0$ opens up a new gapless
asymptotic direction along the vector multiplet \cite{HKY}.
Therefore, $\zeta\rightarrow 0$ does not qualify as a ``small" deformation.

For most physics problems, in fact, this subtlety is associated
with asymptotic flat directions. Whenever we encounter a
supersymmetric theory with (partially) continuous spectrum,
we must be very careful. When the asymptotic direction
is gapped, say at $E=E_{\rm gap}$, this is more of a nuisance than
a problem. The path-integral for the index can yield a non-integral
expression with an additive piece of type
$$\sim e^{-\beta E_{\rm gap}} \ , $$
which, since we cannot take $\beta\rightarrow \infty$ in
actual path-integral, can compromise the computation of the
integral index. Nevertheless, a deformation that scales up
$E_{\rm gap}\rightarrow a\cdot E_{\rm gap}$ with
$a\rightarrow +\infty$ allows us to isolate
the integral index relatively easily. On the other hand,
if one finds an asymptotic direction with $E_{\rm gap}=0$,
as in $\zeta=0$ case discussed above, the subtlety becomes
more pronounced.

Well-known examples of the latter are the nonlinear sigma
models (NLSM's) with a noncompact target manifold. While
the index itself should be a well-defined and integral
object, path-integral computation or alternatively
naive application of Atiyah-Singer index theorem
will usually compute the $\beta \rightarrow 0$ limit
of the trace formula, often denoted as $\cI_{\rm bulk}$, which
is typically non-integral.
To compute the true index, one must also figure out the difference
of the two limits, at $\beta\rightarrow 0$ and $\beta \rightarrow \infty$,
$$
\delta\cI
\equiv
\lim_{\beta\rightarrow \infty}{\rm tr}\left[ (-1)^{F}\cdots \,   e^{-\beta H}\right]-
\lim_{\beta\rightarrow 0}{\rm tr}\left[ (-1)^{F}\cdots \,   e^{-\beta H}\right]  \ .
$$
The only known systematic computation of this so-called
``defect" term or the boundary term, $\delta\cI$,
is for the Atiyah-Patodi-Singer boundary condition,\footnote{For
an accessible review for physicists, see Ref.~\cite{Eguchi:1980jx}.}
which is {\it not} in general equivalent to the physical
$L^2$ boundary condition for the original problem.

A popular alternative that can produce an integral expression,
at the risk of modifying the asymptotic dynamics, is to introduce
chemical potentials \cite{AlvarezGaume:1986nm}. The index, upon such
a modification to the dynamics, is then given as
\bea\label{1.1}
\Omega({\bf y},x;\beta)\equiv {\rm tr}
\left[ (-1)^{F} {\bf y}^{R+\cdots} x^{G_F} e^{-\beta H}\right] \ ,
\eea
where $G_F$ denotes the flavor symmetry generators collectively and
${\bf y}$ is distinguished from the rest of the chemical potentials $x$,
as its exponent involves an R-symmetry generator $R$.
Such a twisted partition function is well-defined and may produce an integral
expression, as long as there is at least one supercharge $Q$ that commutes
with $R+\cdots$ and obeys $Q^2=H$. In case no such
supercharge exists, we simply drop the ${\bf y}$ term. It should be emphasized
that the flavor chemical potentials, $x$, are not as innocuous as
the expression~\eqref{1.1} might lead one to think.
When $|x|\neq 1$, they generate a confining potential to the matter dynamics
which, of course, is not in general ``small," and hence, one needs to be
very careful about them.

Desirably, one must honestly compute
\bea\label{WI}
\cI({\bf y})\equiv \lim_{\beta\rightarrow \infty}{\rm tr}
\left[ (-1)^{F} {\bf y}^{R+\cdots} e^{-\beta H}\right] \ ,
\eea
where the trace is taken over physically acceptable states,
including $L^2$ normalizable bound states.
In this note, we will take much care to distinguish $\Omega$,
the twisted partition function with flavor chemical potentials,
from $\cI$, the Witten index. For theories with {\it compact} dynamics, or
equivalently with fully discrete spectra, the naive deformation arguments
work well and we have
\bea\label{easy}
\cI({\bf y})=\Omega({\bf y},x;\beta) \ .
\eea
In particular, $\beta$- and $x$-{\it independence} of
the right hand side can be shown rigorously for compact
theories \cite{HKY}.
In the presence of {\it gapless asymptotic} directions, however, the
relation (\ref{easy}) is no longer true,
\bea\label{noteasy}
\cI({\bf y})\neq \Omega({\bf y},x;\beta)\ ,
\eea
and there is, {\it a priori}, no
reason to expect why the true Witten index $\cI$ can be
extracted from $\Omega$ in any straightforward manner.

For one thing, $x$-dependence of $\Omega$ can be seen
to survive generically and the desired $x\rightarrow 1$ limit
of $\Omega$ is often divergent. One may still obtain certain integral
information out of $\Omega$ via its Taylor expansion in $x$ or in $1/x$.
While it is tempting to attribute the $x^0$ part of such an expansion
to the flavor-neutral sector of the original index problem, this
can be seen to fail miserably in some of the simplest examples.
For another, when not all of the asymptotic directions in question can
be controlled by $G_F$, one typically finds fractional expressions
for $\Omega$, despite the naive integrality argument. This
can be understood from the fact that in such cases the
path-integral computation of $\Omega$ takes $\beta\rightarrow 0$
limit effectively, leading only to the bulk part of the index.

The main purpose of this note is to explore these two quantities,
the twisted partition function $\Omega$ and the Witten index $\cI$,
for theories with gapless asymptotic directions.
We will find that, indeed, the $L^2$ boundary condition is so
different from the one introduced by ${\bf y}$ and $x$,
and that one computation does not translate to the other in any
obvious manner. One folklore in the community is that deformations
due to chemical potentials, say $x$, will not matter much for the states
that are neutral under the relevant symmetry generator. This, upon
closer inspection, can be seen to have no justifiable ground, however. The
argument may work for classical ground states, completely localized at
fixed points, but quantum states have a wavefunction spread.
An argument of this type may be defensible if we wish to infer
the spectrum at a typical value of $x$ from the limit of $\Omega$
at $\log|x|\rightarrow \pm\infty$, whereby localized
ground state wavefunctions will approach classical ones,
but {\it definitely not} for the spectrum at $\log|x|\rightarrow 0$,
where a wavefunction can easily become nonnormalizable.

Nevertheless, we will note  examples where
some information about $\cI$ is embedded in $\Omega$ in
a very subtle manner whereby one can, {\it a posteriori},
recover $\cI$ from $\Omega$. In one set of examples, we
find that various cohomologies associated with different boundary conditions
can be captured by various flavor-neutral sectors in $\Omega$.
Note that the procedure to isolate a flavor-neutral part is far from being
unique and one is thus led to a multitude of cohomology types.
Surprisingly, the $L^2$ cohomology is
embedded in those flavor singlets, as a common intersection
of two mutually different answers.
In another set of examples, $\Omega$ proves to be fractional rather
than integral, yet $\Omega$ can be understood as a linear
combination of integral $\cI$'s for several related theories.
This allows a simple inversion formula from rational $\Omega$'s
in favor of integral $\cI$'s.
So far, we have no reason to believe that such a recovery of
$L^2$ Witten indices from twisted partition functions is
guaranteed, nor do we see a universal routine for our examples.
Nevertheless, the mere fact that $\cI$'s are somehow embedded in
$\Omega$'s is very suggestive.

This note is organized as follows. Section~\ref{s2} will summarize
the results from Hori-Kim-Yi (HKY)~\cite{HKY}, with some more technical
material relegated to Appendix~\ref{JKR}. Section~\ref{s3} will consider
several classes of theories with a noncompact moduli space from chiral
multiplets, and explore how $\Omega$'s and $\cI$'s are related, with
$\CN=4$ and $\CN=8$ supersymmetries. Here we will see how in general
$\Omega$ fails to contain $\cI$ information properly, or in other words
how there is no obvious way to extract $\cI$ from $\Omega$.
Nevertheless,  we will encounter interesting examples where various flavor expansions
of $\Omega$ seemingly reproduce alternate, albeit physically irrelevant,
cohomologies and how in such examples known $\cI$ is extractable, {\it a posteriori}.
However, we caution the reader that these examples should be taken as anecdotal
rather than exemplary.  Section~\ref{pure} will move on to
more subtle cases with a gapless asymptotic direction in vector
multiplets, which carry fractional $\Omega$'s in the end. Pure
Yang-Mills quantum mechanics are prototypes, and we will provide
detailed relationships between $\Omega$'s and $\cI$'s for this class
of theories, which will lead to new computations and predictions
about threshold bound states for $\CN=4,8,16$ quantum mechanics
for simple gauge groups beyond $SU(N)$.  Motivated by
this, in Section~\ref{quiver}, we turn to the notion of the rational invariant,
which has appeared universally in various wall-crossing
formulae~\cite{KS, JS, GMN1, GMN2, Manschot:2010qz,Manschot:2011xc,Kim:2011sc,Sen:2011aa}.
There, we will also explain how such a notion can help us to count
the very subtle threshold bound states in nonprimitive quiver theories.

\section{Localization}\label{s2}

Quantum mechanical GLSM's with $\cN=4$ supersymmetries can be
obtained by dimensionally reducing $d=4$ $\cN=1$ or $d=2$ $\cN=(2,2)$
systems \cite{GLSM}. Such theories can be characterized by the gauge group $G$ of rank
$r$, the representations $\cR_a$ for the matter fields $\Phi_{a=1, \cdots, A}$,
the Fayet-Iliopoulos (FI) constants $\zeta$ (one for each gauge group
factor with a $U(1)$ sector), and the superpotential. For $d=2$ $\CN=(2,2)$
theories, the FI constant is naturally paired with a $\theta$-angle and becomes
complex. For $d=1$ counterpart, however, physics is independent of
$\theta$ and $\zeta$ can effectively be thought of as a real parameter again,
so that the physics may change discontinuously across the $\zeta=0$ wall,
leading to the wall-crossing phenomena. For the localization computation of
Witten index, which is the main focus of this work, the
superpotential is constrained by $R$-charges, $R_a$,
and other flavor charges, $F_a$, but is assumed generic otherwise.

For an $\cN=4$ GLSM quantum mechanics with hamiltonian
$H$, we are interested in the twisted partition function \cite{HKY,Kim:2011sc,Gaiotto:2010be},
\beq\label{witten-index}
\Omega({\bf y},x;\beta) ={\rm tr}
\left[ (-1)^{2J_3} \bold y^{2J_3 +R} x^{G_F} e^{-\beta H}\right] \ ,
\eeq
where $J_3$ and $R$, respectively, denote the third component of the $SU(2)_R$
generators and the $U(1)_R$ generator, and the flavor symmetry generators
are collectively denoted by $G_F$, with their associated chemical potentials
$x$.\footnote{Normalization of $R$ is such that the superpotential must have
charge $2$ to be consistent with this $R$-symmetry.}
We will sometimes label this object $\Omega$ also by the FI constants,
$\zeta$, to emphasize that it may only be piece-wise constant
in the FI-constant space, possibly exhibiting a wall-crossing behavior.
In this section, a self-contained explanation will be given of
how one can compute $\Omega$ for a simple case. We provide in
Appendix~\ref{A} some technical details needed to deal with a general case
and refer the readers to Ref.~\cite{HKY} for the derivation.\footnote{See also Refs.~\cite{Cordova,Hwang} for related discussions.}

Let us recall that, when the low energy theory is entirely
geometric and compact, this is supposed to capture the following
Hirzebruch-type index \cite{Lee:2012naa},\footnote{This expression differs from
the one used in mathematics literature by the choice of the variable ${\bf y}$ that
encodes the grading, and has an additional overall factor $(-{\bf y})^{-d}$.}
\begin{eqnarray}
\cI_{\rm Hirzebruch}({\bf y})= \sum_{p=0}^d\sum_{q=0}^d
(-1)^{p+q-d}{\bf y}^{2p-d} h^{(p,q)}\ ,
\end{eqnarray}
of the target manifold, where $h^{(p,q)}$ are the Hodge numbers
and $d$ is the complex dimension. For such theories, $\Omega({\bf y},x;\beta)$
reduces to $\cI_{\rm Hirzebruch} ({\bf y})$ and is in particular independent
of $x$ and $\beta$. The overall sign is shifted, relative to
the standard mathematics convention, so that the middle cohomology contributes
with $+$ sign; this feature is built into the localization formulae we present below
in the sign convention for the fermion one-loop determinants.

Via supersymmetric localization,  the path integral for the twisted
partition function leads to~\cite{HKY}
\beq\label{loc}
\Omega^\zeta = \frac{1}{|W|}{\text{JK-Res}}_\eta^{\rm internal}\;g(u)\;{\rm d}^r u \:
+\:\frac{1}{|W|}{\text{JK-Res}}_{\eta;\zeta}^{\rm asymptotic}\;g(u)\;{\rm d}^r u \ ,
\eeq
whose JK-residue \cite{JK} operation is summarized below. In the formula,
$W$ is the Weyl group of $G$ and $u$ collectively denotes the zero
modes of the Cartan vector multiplets, each of which takes values in
a cylinder $\IC^*$ of periodicity $2\pi i$. The quantities
on the right hand side require some further explanations.
Let us briefly clarify what this formula means.

Firstly, the integrand $g(u)$ takes the form,
\beq\label{g}
g(u) = g_{\rm vector}(u)\;  g_{\rm matter}(u) \ ,
\eeq
with the two factors coming from the one-loop determinants of
the non-zero modes for the vector multiplets and those for
the matter multiplets, respectively.
The vector multiplets contribute to the integrand as
\beq\label{gv}
g_{\rm vector}(u) = \left(\frac{1}{{2 \; {\rm sinh} (\frac{z}{2})}}\right)^r
 \; \prod\limits_{\alpha \in \Delta_G} \frac{{\rm sinh}
 (-\frac{\alpha \cdot u}{2})}{{\rm sinh} (\frac{\alpha \cdot u-z}{2})} \ ,
\eeq
where $z$ is defined by $e^{{z}/{2}} = \bold y$ and $\Delta_G$
is the set of root vectors of $G$. The matter multiplets in
turn contribute as
\beq\label{gm}
g_{\rm matter}(u) = \prod\limits_{a=1}^A \prod\limits_{\rho \in \cR_a}
\frac{\;{\rm sinh}\left(-\frac{\rho \cdot u + (\frac{R_a}{2}-1)z
+ F_a \cdot \mu}2\right)}{{\rm sinh}\left(\frac{\rho \cdot u
+ \frac{R_a}{2} z + F_a \cdot \mu}2\right)} \ ,
\eeq
with the flavor chemical potentials  $x=e^\mu$,
where the second product for each matter field $\Phi_a$ is over the
weights $\rho$ of the representation $\cR_a$.

For the $\text{JK-Res}$ operation in eq.~\eqref{loc}, one starts out
by selecting an arbitrary vector $\eta$ of length $r$ \cite{Benini:2013nda,Benini:2013xpa,HKY},
the choice of which does not change the final answer as long as $\eta$
cannot be written as a linear combination of less than $r$ charges.
The latter condition is called ``genericity."
One then proceeds to obtain all the co-dimension $r$
singularities of the integrand $g(u)$ in the $u$-space, $(\IC^*)^r$.
The coordinates of such a singularity will be denoted
collectively as $u_*$ and the collection of charges responsible for $u_*$
as $\bold Q_{u_*}$.
The $\text{JK-Res}$ operation is then given as the summation
over the singularities of the corresponding JK residues, of which
definition we turn to now.
One should not forget that there could be contributions from
co-dimension $r-1$ singularities that extend out to
the asymptotic infinity of the $u$-space, to  which we will come back
momentarily.

For notational simplicity, given a singularity
$u_*$, let us perform a constant shift in $u$ variables, so that the singularity is
located at the origin. Upon such a shift, let us collect all the
arguments of the hyperbolic sine functions of the form, $Q_{i_p}\cdot \,u$,
from the denominator of the integrand. Then, for the singularity $u={\bf 0}$ at the origin,
we have $\bold Q_{\bf 0} = \{Q_{i_p}\}$ in the aforementioned notation,
with each charge in $\bold Q_{\bf 0}$ contributing a linear vanishing to
the denominator. Let us now assume that the origin is a non-degenerate
singularity, {i.e.}, the denominator of the integrand gives rise to exactly $r$
colliding hyperplanes
$Q_{i_p} \cdot u =0$, for $p=1, \cdots, r$.
Then, the JK residue at the origin
is defined as the unique linear functional satisfying
the following properties~\cite{JK, SV},
 \beq
\underset{u=\bf 0}{\text{JK-Res}}\,_{\eta:\bold Q_{\bf 0}}\,
\frac{{\rm d}^r u}{\prod_{p=1}^r Q_{i_p} \cdot u} \equiv
  \begin{cases}
    \frac{1}{|{\rm det}  (Q_{i_1} \,\cdots\, Q_{i_r})|}
    & \text{if } \eta \in {\rm Cone}(Q_{i_1}, \cdots, Q_{i_r}) \\
   0       & \text{otherwise} \ ,
  \end{cases}
\eeq
where ``Cone'' denotes the cone given by positive spans
of the specified charge vectors. In the end, this contributes to the internal part of the JK-Res operation,
\beq\label{JKsum1}
{\text{JK-Res}}_\eta^{\rm internal}\; g(u)\;{\rm d}^r u = \sum_{u_*}
\underset{u=u_*}{\text{JK-Res}}\,_{\eta: \bold Q_{u_*}} \,g(u)\, {\rm d}^r u \ ,
\eeq
where the summation is over all the singularities $u_*$ of co-dimension $r$.

However, the fact that the $u$ variables live in $(\IC^*)^r$ implies that
there could be residues associated with co-dimension $r-1$
singularities that extend out to the asymptotic region of
$(\IC^*)^r$. The contributions from these asymptotic singularities are quite
subtle and have been worked out in Ref.~\cite{HKY} in much detail.
The second term in eq.~\eqref{loc} takes those into account as,
\beq\label{JKsum2}
{\text{JK-Res}}_{\eta;\zeta}^{\rm asymptotic}\;g(u)\;{\rm d}^r u=
\sum_{l_*} \underset{l_*\cap \{\infty\}}{\text{JK-Res}}
\,_{\eta: \bold Q_{l_*}\cup\{Q_\infty=-\zeta\}} \,g(u)\, {\rm d}^r u \ ,
\eeq
where the summation is over all the complex lines $l_*$, given
by the intersection of the singularity hyperplanes for $g(u)$,
and the JK residues are taken at the intersections of $l_*$ and
the asymptotic infinity, denoted as $\{\infty\}$. The symbol
${\bold Q}_{l_*}$ represents the collection of charges, whose
associated singular loci contain the line $l_*$ of singularities.
Note that, whenever available, $-\zeta$ is included as if it
is the charge vector associated with the asymptotic pole,
for the purpose of the JK positivity test.

It may happen that $\zeta$ does not exist in the given theory,
on the other hand, in which case the asymptotic contribution
requires a more careful treatment.
However, when $\zeta$ does exist and when
we can find $\eta=\zeta+\delta$ such that it is generic and belongs to the
same chamber\footnote{The $r$ dimensional vector space of charges
may be divided into chambers where the dividing walls are linear
spans of $r-1$ collections of linearly independent charge subsets.}
as $\zeta$ in the charge space, the asymptotic
singularities can be seen to fail the JK-positivity test
automatically. This allows us to simplify the formula to \cite{HKY}
\beq
\Omega^\zeta = \frac{1}{|W|}{\text{JK-Res}}_{\zeta+\delta}^{\rm internal} \; g(u)\; {\rm d}^r u \ .
\eeq
There are also other circumstances when contributions
from the boundary cancel among themselves completely,
regardless of $\eta$. No matter what $\eta$ is chosen,
we have
\beq
\Omega^\zeta =\frac{1}{|W|} {\text{JK-Res}}_{\eta}^{\rm internal} \; g(u)\; {\rm d}^r u \ ,
\eeq
in such cases.
Pure Yang-Mills theories with a simple gauge group,
which carries no $\zeta$ to begin with, belong to this class,
as outlined in Appendix~\ref{B}.

There are further subtleties to be discussed. First, singularities can be
of a degenerate type,
in which case the number of colliding hyperplanes exceeds
the rank $r$ of the gauge group. In fact, this happens very
generically, including bulk of examples in this note.
In such situations, a constructive definition of the
JK residue can be used~\cite{BV, SV}, which, for the case of
non-degenerate singularities, turns out to be equivalent to the above procedure. Examples in this note tend to have highly degenerate
singularities, for which we devote Appendix~\ref{A} to
describe the actual routine we used.

Second, the JK-residue formulation of $\Omega$ can only be trusted
when the relevant charge sets $\bold Q_{u_*}$ are projective.
The projectivity means that the charges in question can be
considered all ``positive" with respect to some ordering in
the charge space, or equivalently, that they all belong to a half-space.
Unfortunately, we may in general, encounter a degenerate singularity
$u_*$ for which $\bold Q_{u_*}$ is non-projective.
In fact, this happens quite frequently in non-Abelian theories.
The derivation outlined in HKY no longer works for such singularities,
and no systematic mechanism to deal with them is currently available.
The simplest way out would be to deform the
pole locations a little so that $u_*$ will split into several projective ones.
This strategy was successfully used in Ref.~\cite{Kim:2015fba},
where the desired results anticipated on physical ground were correctly
reproduced. We would take this strategy as well if needed,
although, in all of our explicit examples, such nonprojective
poles did not carry any iterated residue and were thus harmless.

Finally, although we have assumed $\CN=4$ in phrasing the localization
prescription, other types of theories, as long as $\CN\ge 2$, can be dealt with
in a similar fashion. HKY in particular was phrased for $\CN=2$ to begin with,
while $\CN=8,16$ is a matter of adding more chiral multiplets.
For the latter, we also need to turn on chemical potentials
that are consistent with the right superpotential.

\section{$L^2$ Index vs. Twisted Partition Function with Chemical Potential}\label{s3}

In the previous section, we have reviewed how twisted partition
function is computed in terms of JK residues via localization.
For this to capture the correct Witten index, the dynamics in question
must obey several conditions, most constraining of which is that there are no
gapless asymptotic directions. However, many physically interesting cases
fail to meet this condition.

When the flat direction has a gap, the problem is
relatively innocuous. As mentioned earlier, the twisted partition function
computation typically yields contributions that scale like
$$~ e^{-\beta E_{\rm gap}}\ ,$$
which can be removed by modifying the dynamics such that
$E_{\rm gap}\rightarrow \infty$. This
procedure does not affect the true index of the dynamics
as the above contribution comes from the continuum
sector at $E\ge E_{\rm gap}$. The JK residue formulae
presented in the previous section follows only after
taking such a deformation of the theory for the
purpose of index computation~\cite{HKY}.

On the other hand, when the flat direction has no gap, $E_{\rm gap}=0$,
we must be more careful. The right thing to do is to impose $L^2$ boundary
condition and compute the index, but this is easier said than done.
>From many works in the recent literature, including those that lead
to the JK residue formulae of the previous section,
the ``index" is computed with chemical potentials inserted, which
introduces massive deformations that lift the flat directions.
For the cases where the original dynamics is compact to begin with,
this merely leads to the equivariant version of the index
and allows a fast and universal computation. When the original dynamics
contains gapless asymptotic direction, however, we must
take more care since the infrared regulator thus introduced
may not be consistent with the natural $L^2$ boundary condition.

In this section, we consider how chemical potential deals with a gapless
asymptotic direction from chiral multiplets.
For $d=1$, in the presence of a gapless Coulombic flat direction at
quantum level, introduction of chemical potential is not quite enough
to control the infrared issue. This is because a gauge multiplet in $d=1$
comes with three noncompact scalars at classical level, instead of two as
in $d=2$. Usual index computation leads to fractional
coefficient, signaling contributions from the continuum sector.
Since such cases are qualitatively different from those with a gapless
chiral flat direction, we will explore and deal with them separately
in Section~\ref{pure} and thereafter.

\subsection{Flavor Expansions of Twisted Partition Functions}

Perhaps the simplest yet instructive example is the free
theory with a single chiral multiplet with its bosonic
degrees of freedom parameterizing a single copy of $\C$.
Turning on the flavor chemical potential $x$ associated with
$U(1)$ rotation of the complex plane as the infrared regulator,
the path integral gives
\bea\label{IC}\Omega^\IC= \frac{x^{1/2}{\bf y}^{-1}- x^{-1/2}{\bf y}}{x^{1/2}-x^{-1/2}}\ .
\eea
Depending on the sign of  $\log|x|$, one gets two different
expansions in $1/x$ or $x$. The former, for example, is
\bea \label{IC1} \Omega^\IC= {\bf y}^{-1}+({\bf y}^{-1}-{\bf y})\cdot(x^{-1}+x^{-2}+\cdots)\ ,\eea
while the latter has the same form except $x\rightarrow 1/x$
and ${\bf y}\rightarrow 1/{\bf y}$,
\bea \label{IC2} \Omega^\IC= {\bf y}+({\bf y}-{\bf y}^{-1})\cdot(x+x^2+\cdots)\ . \eea
Note that the infinite towers, apparent in the expansion
with respect to $x$ or to $1/x$, cannot have a physical meaning
in view of the original free theory. They are the artifacts
of the confining potential introduced by the chemical potential.
One could hope that the flavor-neutral states, at the bottom of
each series, make sense physically. The latter suggests a single
$U(1)$ invariant ground state, but note that the respective
$R$-charges disagree.

There exist cohomologies that are seemingly captured by these
results. Recall that the de Rham cohomology, without any
boundary condition, is
\bea
H^n_{\rm dR}(\C)=\left\{
\begin{array}{lr}
0 &\quad n=2\\
0  &\quad n=1 \\
\R &\quad n=0  \end{array}\right.
\eea
while the de Rham cohomology with compact support is
\bea
H^n_{\rm compact\; dR}(\C)=\left\{
\begin{array}{lr}
\R &\quad n=2\\
0  &\quad n=1 \\
0 &\quad n=0\end{array}\right.
\eea
Recalling how the twisted partition function would be
related to the Hirzebruch index if the low energy theory
were compact, we  see that the flavor-neutral part of
eq.~(\ref{IC1}) captures $H^\bullet_{\rm dR}(\C)$, while the other
expansion~\eqref{IC2} reflects $H^\bullet_{\rm compact\; dR}(\C)$.

Emergence of the two different cohomologies, depending on the
sign of $\log|x|$, tells us that something goes wrong when
one tries to turn off the chemical potential $x\rightarrow 1$
for the original gapless dynamics.
Furthermore, if one is interested in actual bound state wavefunction,
it has to be an $L^2$ normalizable harmonic form, and no such
bound state exists for free theory with the target $\C$.
Neither of the above two expansions, or the two related cohomologies,
counts $L^2$ harmonic forms by itself, in particular.

A slightly more informative example, complete with
a gauge multiplet, is an Abelian GLSM with charges
of both signs appearing in matter fields, in the absence of superpotential.
Consider an ${\cal N}=4$ $U(1)$ theory with $N$ $(+1)$-charged chirals
and $K$ $(-1)$-charged chirals. We assign vanishing $R$-charges
for both sets of chirals so that superpotential is disallowed. With this,
the classical moduli space is contractible to either $\CP^{N-1}$ or
$\CP^{K-1}$. As is
customary, let us control the noncompact directions along the
fibers by introducing chemical potentials.
Of $SU(N)\times SU(K)\times U(1)$ flavor symmetry,
it suffices to introduce a single chemical potential $x$ for the
$U(1)$ flavor under which all charged fields has $+1$ charge.
For example, with $N=3$ and $K=2$, we find \cite{HKY}
\begin{eqnarray}
\Omega^{\zeta>0}_{N=3;K=2} &=& \biggl[
(1+2x^2){\bf y}^{-4}+(1-10x-3x^2){\bf y}^{-2}
\cr\cr&&
+(1-4x^2+24x^4-4x^6+x^8)+(-3x^4-10x^6+ x^8){\bf y}^{2}\cr\cr
&&+(2x^6+x^8){\bf y}^{4}
\biggr]/{(1-x^2)^4} \ .
\end{eqnarray}
For general $N$ and $K$, upon expanding $\Omega$ in either $x$ or $1/x$,
we find
\begin{eqnarray}
&&\Omega^{\zeta>0}_{N;K}=\left\{
\begin{array}{l}(-1)^{N+K-1}\left({\bf y}^{1-N-K}+\cdots+ {\bf y}^{N-K-3} +{\bf y}^{N-K-1}\right)+O(x^2) \\ \\
(-1)^{N+K-1}\left({\bf y}^{1+K-N}+{\bf y}^{3+K-N}\cdots+{\bf y}^{N+K-1}\right)+O(1/x^2) \ ,
\end{array}\right.
\end{eqnarray}
and the expansion for $\zeta<0$ case is obtained by exchanging the roles of
$N$ and $K$.

Again, one possible interpretation of these results is that
these two flavor-neutral parts of the twisted partition function
count, respectively, the de Rham cohomology and the compact-support cohomology,
of the moduli space. The de Rham cohomology is homotopy
invariant, so that we only need to know $H^\bullet(\CP^{N-1})$ for
$\zeta>0$,
\bea
H^{n}_{\rm dR}(\zeta>0)=\left\{
\begin{array}{ll}
0 &\quad n=2k,\; N\le k \le N+K-1\\
\R &\quad n=2k,\; k<N\\
0 &\quad n=2k+1 \ ,
\end{array}\right.
\eea
while the de Rham cohomology with compact support is
obtained from this by Poincare duality as
\bea
H^{n}_{\rm compact\; dR}(\zeta>0)=\left\{
\begin{array}{ll}
\R &\quad n=2k,\; K\le k \le N+K-1\\
0 &\quad n=2k,\; k<K\\
0 &\quad n=2k+1 \ .
\end{array}\right.
\eea
The Betti numbers are entirely represented by $h^{(p,p)}$, so
the leading flavor-neutral parts of the two expansions clearly
capture these two sets of cohomologies. As we elaborate below,
however, neither counts $L^2$ harmonic forms.

Exactly why these cohomologies are embedded in $\Omega$'s
is mysterious.  One might be tempted to argue that,
at least for the compact cohomology,
this is plausible since chemical potential deforms the dynamics
grossly at the asymptotic region. The latter is likely to affect
less those flavor-neutral wavefunctions with compact support,
if there exist any such. For ordinary de Rham cohomology without
boundary condition, however, this becomes even less clear. In fact, we
will also find examples, later on, where this appearance of various
cohomologies  is not a universal phenomenon.

\subsection{Recovering $L^2$ Witten Index?}

A priori, the twisted partition function, $\Omega$, cannot be argued to
compute physical Witten index for theories with asymptotically
flat directions, since the chemical potential introduces
a rather drastic modification to the asymptotic dynamics.
Nevertheless, one could hope that at least flavor-neutral
sectors, often argued to be less sensitive to the chemical
potential, might be captured by $\Omega$. Recall that, for
compact theories, one can rigorously argue that the supersymmetric
ground states are all flavor neutral. However, we learned from
the previous examples that this rosy picture is nowhere near
justified. The notion of flavor-neutral part in $\Omega$ may crucially
depend on the chemical potentials, to begin with, and it does not
necessarily agree with what the true Index would have computed.

In these examples, yet, one notices a curious and encouraging
fact: Although neither of the two Laurent expansions gives us
the correct $L^2$ index for flavor-neutral states, the common
intersection of the two expansions does. In other words,
if we keep only the favor-neutral sectors, and then further
restrict to those states that appear in the both expansions,
this common set is exactly in one-to-one correspondence
with the anticipated $L^2$ states. For the example of $\IC$,
this is trivially so, with no $L^2$ harmonic forms on $\IC=\IR^2$
and no common set in the two expansions for small $x$ and for small $1/x$.

$L^2$ harmonic forms $\Psi$ admit
nondegenerate pairing, $$ \int \Psi\wedge \Psi' \ ,$$ and the
spectrum must be invariant under $\Psi \rightarrow *\Psi$.
This is clearly not the case for the cohomologies with
or without compact support. However, if we carve out
the common intersection of the two cohomologies, Poincare dual
to each other, we do obtain a spectrum that is invariant. More
precisely, for asymptotically conical geometry, there is
a mathematical result due to Hausel et al.~\cite{Hausel:2002xg},
which states\footnote{PY is indebted to Edward Witten for
independently suggesting a possibility of $L^2$ cohomologies
realized as intersection of $H_{\rm dR}$ and $H_{\rm compact\;dR}$;
this is effectively the content of the theorem.}
\bea\label{L2}
H^n_{L^2}(M)=
\left\{\begin{array}{ll}
H^n(M,\partial M) &\quad n< d=({\rm dim}_{\IR}M)/2 \\
{\rm Im}\left(H^n(M,\partial M)\rightarrow H^n(M)\right) &\quad n= d \\
H^n(M) &\quad n> d \ .
\end{array}\right.
\eea
With $H^k(M,\partial M)=H_{2d-k}(M)$, we see immediately that
$H_{L^2}$ is mapped to itself under the Poincare
map.

For the second example, $H^n(M)=H^n_{dR}(M)$ and
$H^n(M,\partial M)=H_{2d-n}(M)=H^n_{\rm compact \, dR}(M)$ \cite{Bredon},
so that
\bea
\cI^{\zeta>0}_{N;K}({\bf y})\biggr\vert_{L^2}=
\left\{\begin{array}{ll}
(-1)^{N+K-1}\left({\bf y}^{1+K-N}+\cdots +{\bf y}^{N-K-1}\right)&N>K \\ \\
0 & N\le K \ .
\end{array}\right.
\eea
As we already noted, this $L^2$  Witten index does not coincide
with the flavor neutral part of the Index with chemical potential,
but, nevertheless, can be extracted by inspecting the two
expansions near $x=0$ and near $1/x=0$, and isolating the
common sector thereof.

Encouraged by this, let us consider a slightly more involved
example of $A_k$ type ALE spaces. These are blowup of $\IC^2/Z_{k+1}$
where $Z_k$ is embedded in holomorphic $SU(2)$ isometry of $\IC^2$,
with $k$ two-cycles parameterized by the $k$ blowup parameters.
While these spaces are hyperK\"ahler, there is a simpler
$\CN=4$ GLSM realization with a choice of
preferred complex structure. For this, we start with a
$U(1)^k$ gauge group with $k+2$  chiral multiplets of
gauge charges,
\bea
&(1,~-2,~~1,~~0,&0,~~0,~\dots,~~0) \ , \cr
&(0,~~1,~-2,~~1,&0,~~0,~\dots,~~0) \ , \cr
&\quad\vdots&\quad\quad\vdots \cr
&(0,~\cdots,~~0,~~0,&0,~~1,~-2,~~1) \ ,
\eea
where each line represents the gauge charges of the $k+2$
chirals with respect to a single gauge $U(1)$. We obtain
the resolved $A_k$ ALE space by turning on $k$ positive
FI constants. Since the asymptotic geometry is conical $\IC^2/Z_k$,
flavor chemical potential is needed. Only two $U(1)$ flavor
symmetries are effective, which we can choose to be represented
by the two flavor charge vectors,
\bea
&&(1,~~0,~~0,~~0,~\dots,~~0,~~0,~~1)\ ,\cr
&&(1,~~0,~~0,~~0,~\dots,~~0,~~0,-1) \ .
\eea
Similarly, as in the previous example, let us turn on the
chemical potential $x$ associated with the first of these
two. Computation of the twisted partition function gives
\bea
\Omega({\bf y},x)^{\zeta_i>0}_{A_k}=\frac{(1+x^{k+1}) (x^2{\bf y}^{-2}+{\bf y}^2)+
(k(1+x^{k+3})-(k+2)x^2(1+x^{k-1}))}{(-1+x^2)(-1+x^{k+1})} \ ,
\eea
from which one quickly realizes that
the expansion with respect to $x$ or $1/x$ yields, respectively,
\bea\label{Ak-expansions}
\Omega({\bf y},x)^{\zeta_i>0}_{A_k}&=& k +{\bf y}^{2} + O(x) \ ,\cr\cr
\Omega({\bf y},x)^{\zeta_i>0}_{A_k}&=& {\bf y}^{-2}+ k +O(1/x) \ ,
\eea
which suggest, following the same line of thought as above,
the common coefficient to ${\bf y}^0$, i.e. $k$, as the true Witten index.
This indeed reproduces the well-known
$L^2$ spectrum of $M_{A_k}$: It is not difficult
to see that the flavor-neutral
sectors of these two expansions capture the two different
cohomologies of $A_k$ space,
\bea
H^{n}_{\rm dR}(M_{A_k})=H^n(M_{A_k})=\left\{
\begin{array}{ll}
\R &\quad n=0,\\
\R^k &\quad n=2,\\
0 &\quad {\rm otherwise} \ ,
\end{array}\right.
\eea
and
\bea
H^{n}_{\rm compact\; dR}(M_{A_k})=H^n(M_{A_k},\partial M_{A_k})=\left\{
\begin{array}{ll}
\R^k &\quad n=2,\\
\R &\quad n=4,\\
0 &\quad {\rm otherwise} \ ,
\end{array}\right.
\eea
respectively, while the $L^2$ cohomology is,
either from eq.~(\ref{L2}) or from past experiences with
these spaces in string theory,
\bea
H^{n}_{L^2}(M_{A_k})=\left\{
\begin{array}{ll}
\R^k &\quad n=2,\\
0 &\quad {\rm otherwise} \ .
\end{array}\right.
\eea
Again, this cohomology with physical boundary condition is
embedded in $\Omega$ as the common sector between
the two expansions~\eqref{Ak-expansions} with respect to $x$ and $1/x$,
just as in the previous rank-one example.

However, note that we have forgotten about the other $U(1)$
flavor symmetries, represented by the chiral charges $(1,0,0,\dots,0,-1)$. Using its chemical
potential instead, say, $x'$, and taking the limits thereof, we
find a different pattern,
\bea\label{Ak}
\Omega({\bf y},x')^{\zeta_i>0}_{A_k}&=& k+1 +O(x') = k+1  +O(1/x') \ .
\eea
Following the same prescription as the case of $x$ would
lead to a wrong answer of $k+1$. The choice of chemical
potentials to be used as infrared regulator is thus also important,
if one wishes to learn about $L^2$ state counting. Turning on both
$x$ and $x'$, this subtlety translates to an order of limit issue.
This type of subtleties is quite generic, making the
$L^2$ index extraction out of $\Omega$ more of an art. For this
particular example, the symmetry associated with $x$ is the one
$U(1)$ isometry that is present for generic $A_n$ manifold, and
the one with $x'$ is accidental feature of using $\CN=4$ GLSM
realization. Whether this has a particular meaning in the
above context is not clear.

Furthermore, these examples so far can hardly be argued to be
representative of most general gauged quantum mechanics
with gapless asymptotic directions. The former are all Abelian,
with rather simple gauge charges. Yet, the manner in which the
$L^2$ cohomology information is embedded in $\Omega$ is rather
intriguing and deserves to be explored further. Some of the
obvious questions that need to be answered next are,

\noindent
{\it Q1: Is there any further simplification if we demand
larger supersymmetry?}

\noindent
{\it Q2:  Does such a simple relation between true index
$\cI_{L^2}({\bf y})$ and twisted partition function $\Omega({\bf y},x)$
persist in more general theories? }

\noindent
It turns out that a useful laboratory for addressing these questions
is the $\CN=8$ system of ADHM, to which we turn next.

\subsection{$\CN= 8$ and $U(1)$ ADHM}

Let us consider $\CN=8$ GLSM in general. Each
hypermultiplet consists of a pair of $\CN=4$ chirals,
$(H_1^f, H_2^f)$, in mutually conjugate gauge representations
while the $\CN=4$ vector multiplet is augmented by an adjoint
chiral multiplet $\Phi$ \cite{Witten:1995gx}. The complex part of the triplet
of $D$-terms can be regarded as superpotential that has
the universal form
$$\sum_f H_1^f\Phi H_2^f \ . $$
In addition to the $U(1)_R$ of $\CN=4$ subalgebra, which
we can choose such that
$$R_\Phi=2\ , \quad   R_{H_1^f}=R_{H_2^f}=0\ , $$
there are two different types of $\CN=4$ flavor symmetries.
One is the universal symmetry, $\tilde R$, that rotates $\Phi$ as well
as all the $H$'s with the charges
$$\tilde R_\Phi=-2\ , \quad  \tilde R_{H_1^f}=\tilde R_{H_2^f}=1 \ , $$
which appears as a flavor symmetry in $\CN=4$ sense but really
is part of $\CN=8$ superalgebra. Let us denote
its chemical potential as $\z=e^{w/2}$.
The other type of symmetries, $\tilde F^{f}$, only rotates the $f$-th
hypermultiplet, $(H_1^f, H_2^f)$, with
$$(\tilde F_{H_1^f}^f,\tilde F_{H_2^f}^f)=(1,-1)\ , $$
while $\tilde F_{H_i^{f'\neq f}}^f=0$ and $\tilde F_\Phi=0$.
Thus, they are truly a flavor symmetry of $\CN=8$ theories.

Denoting by $\tilde \mu$ the chemical potential for $\tilde F$'s
collectively, note that $\Omega$ is always an even
function of $\tilde \mu$. In the integrand $g$ of the
residue formulae, contributions from $H^f_1$ and $H^f_2$
are identical to each other except $\tilde\mu^f\rightarrow -\tilde \mu^f$.
This means that the two expansions of
$\Omega({\bf y}, \z, \tilde x)$
under $\tilde x=e^{\tilde \mu}$ and under $1/\tilde x=e^{-\tilde\mu}$
are necessarily identical and in particular share the identical
$\tilde F$-neutral sector, captured by
\bea
\tilde \Omega({\bf y}, \z)\equiv\Omega({\bf y}, \z, \tilde x)\biggr\vert_{\tilde x\; \rm flavor \;neutral\; sector} \ .
\eea
Although we will mostly need to turn on the flavor chemical
potentials associated with $U(1)$ factors only, rather than the Cartan
part of a non-Abelian flavor symmetry, this observation applies to
all such $\CN=8$ flavor symmetries. If the non-Abelian flavor
chemical potentials are used, however, we should be more
careful in extracting the singlet part by carefully organizing
the expression via characters.

For $\tilde R$, however, there is no reason to expect the same
simplification despite the higher supersymmetry. As with generic
flavor chemical potential $x$ of $\CN=4$ cases, expansions
about $\z=0$ and $1/\z=0$ would in general disagree. If the
same pattern persists, we might expect to find the result to
encode $\cI$ of the low energy manifold from
careful inspection of the limiting expressions,
\bea
\tilde \Omega({\bf y}, \z)\biggr\vert_{\z^{\pm 1}\rightarrow 0} \ .
\eea
Note that the choice of ${\bf y}$ as the preferred $R$-symmetry chemical
potential that carries the cohomology grading is in principle
ambiguous, since it is tied to the choice of $\CN=4$ out of $\CN=8$.
On the other hand, this  reflects the choice of the complex structure
on the hyperK\"ahler vacuum manifold, and should not matter for
the shape of the Hodge diamond.

For an illustration, consider ADHM quantum
mechanics with $U(k)$ gauge group and $U(N)$ flavor group,
i.e., the D0-D4 system with $k$ D0's and $N$ D4's. For these
theories, FI constants are allowed and lift the would-be
Coulombic vacua, so the gapless asymptotic flat
direction arises entirely from hypermultiplets only, and the
twisted partition function can compute a proper integral index. Furthermore,
$\CN=8$ implies that the FI constants can be thought of as a
triplet, so that $\zeta=0$ does not divide FI constant space
into two domains. No wall-crossing can occur, as is also
evident from the explicit computations. The classical moduli
space will be of the form $M_{k;N}\times \IR^4$ where the latter is the
decoupled trace part of the adjoint hypermultiplet while the
former is the moduli space of centered $k$ instantons of $U(N)$
gauge theory. Will the same procedure as in the previous examples
compute $L^2$ Index of $M_{k;N}$?

Theories for a single D0  are the simplest. For example,
$k=1$ and $N=1$ gives
\bea
\Omega^{k=1;N=1}_{\rm ADHM}({\bf y},\z,\tilde x)=\frac{({\bf y}\z^{-1} \tilde x^{1/2}-{\bf y}^{-1}\z
\tilde x^{-1/2})( {\bf y}^{-1}\z \tilde x^{1/2}-{\bf y}\z^{-1} \tilde
x^{-1/2})}{ (\z \tilde x^{1/2} -\z^{-1}\tilde x^{-1/2})(\z^{-1}
\tilde x^{1/2}-\z \tilde x^{-1/2})} \ ,
\eea
where $\tilde x$ is from the flavor $U(1)$ of the adjoint hypermultiplet.
This happens to equal that of a single free hypermultiplet,\footnote{
Expansions in $1/\tilde x$ and in $\tilde x$ produce
\bea \label{IH1} 1 +\left[ (\z^{-1}-\z)^2-(\z/{\bf y}-{\bf y}/\z)^2\right]\tilde x^{-1}+O(\tilde x^{-2})\ , \eea
and
\bea \label{IH1} 1 +\left[ (\z^{-1}-\z)^2-(\z/{\bf y}-{\bf y}/\z)^2\right]\tilde x+O(\tilde x^{2})\ , \eea
so that, unlike the case of a free chiral multiplet, the presence
of a free hypermultiplet gives 1 in the end in our expansions
relative to flavor and $\tilde R$ chemical potential. In the
context of ADHM, the trace part of the adjoint hypermultiplet
decouples and represents the center of mass motion of the
instantons, yet, as far as this procedure goes, is pretty harmless.} which
in the past encouraged a single bound state interpretation.
However, the matter is a bit more subtle, as we presently explore.
Taking flavor neutral sector by expanding in either $\tilde x$
or $1/\tilde x$ and setting either to zero gives $\tilde\Omega({\bf y},\z)= 1$,
which we are encouraged to interpret as
\bea
\cI^{k=1;N=1}_{\rm ADHM}\biggr\vert_{L^2}= 1 \ .
\eea
More generally we find
\bea\label{k=1}
\tilde \Omega^{k=1;N}_{\rm ADHM}({\bf y},\z\rightarrow 0) &=& 1+{\bf y}^2+\cdots {\bf y}^{2N-2} \ ,  \cr\cr
\tilde \Omega^{k=1;N}_{\rm ADHM}({\bf y},\z\rightarrow \infty) &=& 1+{\bf y}^{-2}+\cdots {\bf y}^{2-2N} \ ,
\eea
which contain a single common sector corresponding to $1\cdot {\bf y}^0$
but otherwise, again, disagree with each other.

Recall that the  ADHM moduli space has the form
$M_{\rm ADHM}^{k=1;N}\times \IR^4$, where the latter factor
is the free center of mass part and $M_{\rm ADHM}^{k=1;N}$
is $4(N-1)$ dimensional manifold that is contractible to
$\IP^{N-1}$. Thus, we have
$$H^\bullet(M_{\rm ADHM}^{k=1;N})=H^\bullet(\IP^{N-1}) \ ,$$
while
$$H^\bullet(M_{\rm ADHM}^{k=1;N},\partial M_{\rm ADHM}^{k=1;N})$$
is the Poincare dual thereof. Thus, $L^2$ cohomology of $M_{\rm ADHM}^{k=1;N}$,
again using eq.~(\ref{L2}), is
\bea
H^{n=2N-2}_{L^2}(M_{\rm ADHM}^{k=1;N})=\IR \ ,
\eea
and vanishes for other $n$'s, such that
\bea
\label{k=1L2} \cI_{\rm ADHM}^{k=1;N}\biggr\vert_{L^2}=1 \ .
\eea
We again find that the common sector of the two expansions
of $\tilde \Omega^{k=1;N}_{\rm ADHM}({\bf y},\z)$ captures precisely
the same answer as this.

Note that this result is qualitatively on par with
$\CN=4$ examples and show no real advantage
due to larger supersymmetry. There is some simplification
in that  the two expansions with respect to $\tilde x$ and
$1/\tilde x$ coincide, where $\tilde x$ is associated
with the flavor symmetry in $\CN=8$ sense. However, the expansions
for $\z$ and $1/\z$ do not agree with each other as in $\CN=4$
examples, and
true Index emerges only after taking the common sector appearing
in both expansions. This tells us that $\CN=8$ does not buy us
too much; Existing claims about bound states based on twisted
partition function with chemical potential  must be thus taken
with a large grain of salt.

\subsection{General ADHM and Multi-Particle Contributions}\label{s3.4}

Turning to $k>1$, however, we see that such a pattern does not
persist. Repeating the same procedure, and
considering $N=1$ only, we find
\bea\label{12357}
\tilde\Omega^{k;N=1}_{\rm ADHM}({\bf y},\z^{\pm 1}\rightarrow 0)
= 1,2,3,5,7,\cdots \qquad \hbox{for } k=1,2,3,4,5,\cdots \ .
\eea
With this new example, we see immediately that the previous
instances of various flavor expansions of $\Omega$ or $\tilde \Omega$
giving some version of cohomology were accidental. The moduli space
$M^{k;N=1}_{\rm ADHM}$ is $4k-4$ real-dimensional, and its cohomology
must at least have $H_0(M)$ and $H^{4k-4}(M,\partial M)$ nontrivial.
Neither of such cohomologies is reflected in the above expansions.

One could be in principle  encouraged by the mere integers
unambiguously showing up in eq.~\eqref{12357}, but the numbers
are in a clear conflict with M-theory predictions of unique $L^2$
ground state. In fact, the above numbers in $\tilde\Omega^{k;N=1}_{\rm ADHM}$
are nothing but the number of partitions for $k$, suggesting that
each asymptotic sector with bound states of $k'<k$ D0 branes
also contribute 1 each. One way to phrase this suggestion
starts with the generating function \cite{Kim:2011mv},
\bea
\tilde G^{(N=1)}_{\rm ADHM}(q;{\bf y},\z)
=\sum_{k=1}^\infty q^k\tilde \Omega^{k;N=1}_{\rm ADHM}({\bf y},\z) \ ,
\eea
of which Plethystic logarithm gives
\bea
G^{(N=1)}_{\rm ADHM}(q;{\bf y},\z)
=\sum_{n=1}^\infty  \frac{\mu(n)}{n}\log\left[ \tilde G^{(N=1)}_{\rm ADHM}(q^n;{\bf y}^n,\z^n)\right] \ ,
\eea
where the M\"obius function $\mu$ is defined as
$$\mu(n)=\left\{
\begin{array}{ll}
(-1)^p \qquad & {\text{if~}}n{\; \rm is\; a\; product\; of\;} p\ge 0 {\;\rm  distinct\; primes}\ ,\\ \\
0 & {\rm otherwise} \ .
\end{array}\right.
$$
The proposal boils down to the statement that, expanding
$G^{(N=1)}_{\rm ADHM}$ again in terms of $q$,
\bea
G^{(N=1)}_{\rm ADHM}(q;{\bf y},\z)
= \sum_{k=1}^\infty q^k\Lambda^{k;N=1}_{\rm ADHM}({\bf y},\z) \ ,
\eea
gives us a new set of index-like quantities
$\Lambda^{k;N=1}_{\rm ADHM}({\bf y},\z)$ that carries the information of
one-particle bound state. Since $\tilde \Omega^{k;N=1}_{\rm ADHM}$
happen to be mere numbers, the same is true of
$\Lambda_{\rm ADHM}^{k;N=1}$, and one finds
\bea
\Lambda^{k;N=1}_{\rm ADHM}({\bf y},\z)=1 \qquad \hbox{for all } k\ge 1 \ ,
\eea
which is taken to imply
\bea
\cI^{k;N=1}_{\rm ADHM}\biggr\vert_{L^2}=1 \qquad \hbox{for all } k\ge 1 \ ,
\eea
as anticipated from M-theory.
Of course this does not really
explain why each partition of $k$ contributes exactly
unit; it  merely offers an observation of
a natural relationship between $\cI^{k;N=1}_{\rm ADHM}\vert_{L^2}$'s and
$\Omega^{(k'\le k;N=1)}_{\rm ADHM}$'s.  It also tells us that extraction of
$\cI_{L^2}$ index from $\Omega$, if any such universal
routine does exist, could be a bit more involved
than what we have seen in the earlier Abelian examples.

Furthermore, ADHM for $U(N)$ instantons with $N>1$ offers even more
challenges, as the end results cannot be encoded in a naive
generalization of  the $N=1$ sequence. Let us list some of
the results for $\tilde \Omega^{k;N}_{\rm ADHM}$:

\bea
\tilde\Omega^{k=1;N=2}_{\rm ADHM} ({\bf y},{\z^{\pm 1}\rightarrow 0}) &=&  1 + {\bf y}^{\pm 2}\ , \cr\cr
\tilde\Omega^{k=2;N=2}_{\rm ADHM} ({\bf y}, {\z^{\pm 1}\rightarrow 0}) &=&   3 + 2{\bf y}^{\pm 2}\ ,\cr\cr
\tilde\Omega^{k=3;N=2}_{\rm ADHM} ({\bf y}, {\z^{\pm 1}\rightarrow 0}) &=&   5  + 5{\bf y}^{\pm 2}\ ,\cr\cr
\tilde\Omega^{k=4;N=2}_{\rm ADHM} ({\bf y}, {\z^{\pm 1}\rightarrow 0}) &=&  10 + 9 {\bf y}^{\pm 2} + {\bf y}^{\pm 4}\ ,
\eea

\bea
\tilde\Omega^{k=1;N=3}_{\rm ADHM} ({\bf y}, {\z^{\pm 1}\rightarrow 0}) &=& 1+ {\bf y}^{\pm 2} + {\bf y}^{\pm 4}\ ,\cr\cr
\tilde\Omega^{k=2;N=3}_{\rm ADHM} ({\bf y}, {\z^{\pm 1}\rightarrow 0}) &=&  3  + 3{\bf y}^{\pm 2} + {\bf y}^{\pm 4}\ ,\cr\cr
\tilde\Omega^{k=3;N=3}_{\rm ADHM} ({\bf y}, {\z^{\pm 1}\rightarrow 0}) &=&  6 + 7{\bf y}^{\pm 2} + 7{\bf y}^{\pm 4} + 2{\bf y}^{\pm 6}\ ,
\eea

\bea
\tilde\Omega^{k=1;N=4}_{\rm ADHM} ({\bf y}, {\z^{\pm 1}\rightarrow 0}) &=& 1 + {\bf y}^{\pm 2}
+ {\bf y}^{\pm 4} + {\bf y}^{\pm 6}\ , \cr\cr
\tilde\Omega^{k=2;N=4}_{\rm ADHM} ({\bf y}, {\z^{\pm 1}\rightarrow 0})
&=& 3 + 3{\bf y}^{\pm 2} + 4{\bf y}^{\pm 4} + 3{\bf y}^{\pm 6} + {\bf y}^{\pm 8}\ .
\eea
The common flavor neutral parts are represented by single integers,
in each case of $(k,N)$ ADHM, which are $1,3,5,10,\dots$ for $N=2$; $1,3,6,\dots$ for $N=3$;
$1,3,\dots $ for $N=4$. The M-theory prediction is again a unique
$L^2$ bound state for all of these ADHM theories, so for $k>1$,
multi-particle interpretation is again needed for this to make sense,
although,  unlike $N=1$ example,  simple enumeration of partitions
of $k$ does not suffice.

To summarize, we have seen that physical $L^2$ boundary condition cannot
be mimicked by introduction of chemical potentials as infrared regulators.
In some cases, the latter twisted partition functions capture some version
of cohomologies, depending on which Taylor expansion is taken, whose common
flavor singlet parts turned out to equal the desired $L^2$ spectra.
However, as we have seen in ADHM examples, this particular
pattern proves to be accidental; Even though example by example $L^2$
spectrum seems to be embedded in $\Omega$ in one way or another, no clear
and general dictionary between $\Omega$ and $\cI$ appears to exist.
The situation does not improve upon imposing larger $\CN=8$ supersymmetry, and
one finds again ambiguous flavor-singlet contents typically.
Nevertheless, the mere fact that  $L^2$ spectrum is embedded in $\Omega$
in some indirect manner is by itself pretty miraculous,  given
how callously the chemical potential deals with the subtle infrared issues.
Relationships between  $\Omega$ and $\cI$  we have uncovered
might hint at a more universal relationship for noncompact dynamics.
Further investigation along this line is  desired.

\section{Bulk, Boundary, and Pure Yang-Mills}\label{pure}

So far we have explored examples with gapless asymptotic directions
coming from matter multiplets. In these examples, chemical potentials can
lift all such directions, resulting in integral $\Omega$'s, which
nevertheless must be interpreted with much care. When gapless
directions come from vector multiplets, however, there arise further
issues. For instance, if no FI constants can be turned on,
there would be rank-many Coulombic flat directions that
cannot be lifted by chemical potential.
With such theories, suppose that we still take
the final expressions of type (\ref{loc}) at face value.
What have we computed?

Note that the formula (\ref{loc}) is independent
of $\beta$. At least naively, this is hardly surprising, as
Index counts only $H=0$ states. However, the truth of the matter
is more complicated. For theories with asymptotically
flat directions, gapped or gapless, the continuum
sector with $E>E_{\rm gap}$ can contribute to the path integral
in such a way that the twisted partition function acquires
$\beta$-dependence. As we noted already, we first try to
deform the theory such that $\beta E_{\rm gap}\rightarrow\infty$
to remove such a continuum contribution whenever possible.
On the other hand, this is not possible for gapless asymptotic directions,
and we may acquire contributions from continuum that touches $E=0$
and $\Omega$ would be $\beta$-dependent. Such directions from
the chiral sector can be controlled by introducing flavor chemical
potentials, which lead to both problems and some promises as
described already. For such directions from a vector multiplet, however,
even this is not possible. We must thus expect $\beta$-dependence
of $\Omega$ in general, and the formula~(\ref{loc}), being independent of
$\beta$, must be a certain limit of $\Omega(\beta)$ for such gapless theories.

It would have been nice if the formula (\ref{loc}) had computed the
$\beta\rightarrow \infty$ limit of $\Omega$, with respect to
some boundary condition. Since we do encounter non-integral $\Omega$, however,
this is not possible; True Index, regardless of boundary condition,
would have to be integral. The obvious answer is then that
we have computed the other limit of the twisted
partition function at $\beta\rightarrow 0$, or equivalently the ``bulk" part of the Index.
For pure Yang-Mills theory, this is easiest to see:
Since the theory is not compact, the only other parameter
that enters the computation is the electric coupling $e^2$ with
the dimension of mass cubed. The only dimensionless combination,
$e^{2/3}\beta$, vanishes in the localization limit of $e^2\rightarrow 0$,
implying that the localization formula~(\ref{loc}) effectively computes
the bulk part of the Index,
$$\cI_{\rm bulk} =\Omega\biggr\vert_{e^{2/3}\beta\rightarrow 0} \ .$$
For simplicity, we will continue to use the notation $\Omega$,
for the twisted partition function computed by eq.~(\ref{loc}).

The bulk part of the Index is often non-integral.
Contribution from the continuum sector must be computed separately,
say, $-\delta\cI$, and subtracted from this to produce the true Index,
$$\cI=\cI_{\rm bulk} +\delta \cI\ .$$
Is there a systematic way to compute $\delta \cI$?
Since $L^2$ is natural for path integral, the boundary contribution
with the $L^2$ condition might carry its own physical
meaning. In the well-known D0 bound state problem \cite{Witten:1995ex,Yi:1997eg,Sethi:1997pa}
decades ago, exactly such an interpretation was found~\cite{Yi:1997eg},
which lead to a method of computing the boundary contribution of a theory
as the bulk contribution of a different, much simpler theory.
We will see later that this behavior is quite prevalent,
and in particular quite generic in the context of BPS quivers
and the associated wall-crossing phenomena.

In this section, we will first concentrate on a simplest
possible class of systems, the $\cN=4,8,16$ pure Yang-Mills quantum mechanics,  where the bulk and the boundary are related.

\subsection{${\cal N}=4,8$ Pure Yang-Mills}

For pure Yang-Mills with a simple gauge group,
the Fayet-Iliopoulos constant is absent, whereby we lose
the trick of $Q_\infty=-\zeta$. Nevertheless, one can still argue that
the residue contribution from poles at infinities of $(\IC^*)^r$
cancel among themselves.\footnote{See Appendix
\ref{B}.} The localization procedure gives, then,
$$\Omega^G({\bf y},x;e^{2/3}\beta\rightarrow 0)
=\frac{1}{|W|}{\text{JK-Res}}_\eta^{\rm internal}\; g_G(u)\;{\rm d}^r u \ ,
$$
where $G$ labels the gauge group. Contributing singularities would in general
include highly degenerate ones, for which we resorted to the constructive
procedure outlined in Appendix~\ref{A}.

It turns out that, for pure $\CN=4$ $G$-gauged quantum mechanics,
this bulk contribution to the index, or the $e^{2/3}\beta\rightarrow 0$ limit of
twisted partition function can be organized into \emph{}a universal formula,
\bea\label{N=4}
\Omega^G_{\CN=4}({\bf y})=
\frac{1}{|W|}\sum'_{w}\frac{1}{{\rm Det}\left({\bf y}^{-1}-{\bf y}\cdot w\right)}\ ,
\eea
where the sum is only over the elliptic Weyl elements and
$|W|$ is the cardinality of the Weyl group itself. An elliptic
Weyl element $w$ is defined by absence of eigenvalue 1.
In other words, in the canonical $r$-dimensional representation
of the Weyl group on the weight lattice,
$${\rm Det}\left(1-w\right)\neq 0\ .$$
For pure ${\cal N}=8$ $G$-gauged quantum mechanics, obtained by adding to the
$\CN=4$ theory an adjoint chiral, we have a flavor chemical potential $x$
of the adjoint to play with as well. With $R=0$ for the adjoint chiral, we
have
\bea\label{N=8}
\Omega^G_{\CN=8}({\bf y},x)=\frac{1}{|W|}\sum'_{w}\frac{1}{{\rm Det}\left({\bf y}^{-1}-{\bf y}\cdot w\right)}\cdot
\frac{{\rm Det}\left({\bf y}^{-1} x^{1/2}-{\bf y} x^{-1/2}\cdot w\right)}{{\rm Det}\left( x^{1/2}- x^{-1/2}\cdot w\right)}\ ,
\eea
where again the sum is over the elliptic Weyl elements of $G$.

Why such a universal and simple formula? This can be motivated
by the statement that $\CN=4,8$ pure Yang-Mills quantum mechanics has no
bound state, i.e., the true $L^2$ index is zero. We know that this is the
case at least for $G=SU(N)$. D2/D3-branes multiply-wrapped on
$S^2$ and $S^3$ in noncompact Calabi-Yau two-fold and three-fold are
governed by such dynamics. Judging from how such geometrically engineered
$d=4$ $\CN=2,4$ Yang-Mills field theory behaves, we can pretty much rule
out physical bound states of many identical D-branes wrapping on
$S^2$ and $S^3$. The worldvolume dynamics on such multiply-wrapped
D-brane are $\CN=4,8$ pure $SU(N)$ quantum mechanics, which
tells us the Witten index of these theories must vanish.
If we assume that the latter statement extends to all gauge groups,
we have
\bea
\cI^G_{\cN=4,8}=\Omega^G_{\cN=4,8}(e^{2/3}\beta\rightarrow 0)+\delta{\cI}^G_{{\cN=4,8}}=0\ ,
\eea
where $\delta\cI$ is the contribution associated with the continuum sector.
This implies
\bea
\Omega^G_{\cN=4,8}(e^{2/3}\beta\rightarrow 0)=\cI_{\rm bulk}=-\delta{\cI}^G_{\cN=4,8} \ .
\eea
We have transformed, apparently, a cumbersome problem of
computing the bulk contribution to $\cI$ into an even more
difficult problem of computing continuum contribution $\delta \cI$.
Why is this helpful?

This simple
observation can be made into a computational tool as follows~\cite{Yi:1997eg},
based on the observation that $\delta\cI$ is entirely a boundary contribution.
Since $\delta \cI$ depends only on the dynamics at the asymptotic region,
we need to ask how the latter looks like. The low energy theory in the
Coulomb phase is free with the target being the asymptotic part of
the orbifold,
\bea
\CO(G)=\IR^{3r}/W\;{\rm or}\;\IR^{5r}/W \ ,
\eea
for $\CN=4,8$, respectively, where $W$ is the Weyl group of $G$.
Let us then consider  $\cN=4,8$ $U(1)^r$ gauge theory without matter,
and gauged by the discrete group $W$ acting on the $U(1)$'s. The latter
is a valid quantum theory of its own, and shares the same asymptotic
dynamics as the original Yang-Mill theory.  We will denote this free
theory also by the same symbol $\CO(G)$. Since the interacting
$G$-Yang-Mills theory and the free $\CO(G)$ Abelian theory shares
the same asymptotic dynamics, we conclude that
\bea
\delta{\cI}^G = \delta\cI^{\CO(G)} \ .
\eea
On the other hand, the free orbifold $\CO(G)$ cannot possibly have
a bound state either when $G$ theory does not, since the former can
be considered a limit of the latter, so its bulk
contribution to the Witten index is such that
\bea
\cI^{\CO(G)}_{\rm bulk} =- \delta\cI^{\CO(G)} \ ,
\eea
which finally brings us to
\bea
\Omega^G_{\cN=4,8}(e^{2/3}\beta\rightarrow 0)=-\delta{\cI}^G_{\cN=4,8}=- \delta\cI^{\CO(G)}_{\cN=4,8}
 =\left(\cI^{\CO(G)}_{\cN=4,8}\right)_{\rm bulk}\ .
\eea
The right-most expression can be easily evaluated using the
Heat Kernel regularization, when ${\bf y}=1$ and $x=1$, following the
$SU(2)$ case in Ref.~\cite{Yi:1997eg}, and this gave~\cite{Green:1997tn,Kac:1999av}
\bea
\frac{1}{|W|}\sum'_{w}\frac{1}{{\rm Det}\left(1- w\right)} \ ,
\eea
for all $\cN$.\footnote{$\CN=2$ case also gives the same expression
for $\cI^{O(G)}_{\rm bulk}$ but its relation to $\delta\cI^G_{\CN=2}$ is no longer
justified due to the logarithmic nature of the wavefunctions.}
The expressions in eqs.~\eqref{N=4} and~\eqref{N=8} for $\cN=4,8$ pure
Yang-Mills dynamics are merely equivariant generalizations of this
result. $G=SU(N)$ cases already have ample and independent evidences
that argue against threshold bound states, and the above line
of thinking, combined with our direct localization computation,
further supports the same conclusion. With this class of examples
understood, the mere fact that we recover exactly the same
type of rational structure for other $G$'s involving only the
elliptic Weyl elements, from the brute-force localization computation,
indicates convincingly that threshold bound states are absent for
general $G$ as well.

We will later see, when we turn to $\CN=16$ and also to $\CN=4$
nonprimitive quivers, how this rational structure expands in a
very logical manner in the presence of threshold bound states.
For example, for $\cN=16$ where threshold bound states are
generally expected, $\delta\cI^G_{\CN=16}$ can be seen to receive
contributions from continuum sectors involving partial bound states.
Thus, one ends up with a recursive form of $\cI^G_{\rm bulk}=
\Omega^G_{\cN=16}(e^{2/3}\beta\rightarrow 0)$,
built up from $\cI^{\tilde G}_{\cN=16}$ and $\cI^{\CO(\tilde G)}_{\rm bulk}$
where $\tilde G$'s are subgroups of $G$. We will come back to this story
in Section~\ref{s4.2}.

\subsubsection{Examples}

We have checked the results above against computations following
HKY derivation of $\Omega$ \cite{HKY}. Details of the numerical
computation is often too lengthy to describe, as it can easily
involve thousands of flags and singularities with contributing residues;
We merely state here that {\it the predictions in Eq.~(\ref{N=4}) and in
Eq.~(\ref{N=8}) have been confirmed,}  by such explicit evaluations
of the relevant JK residue formulae, up to rank 5 for $\CN=4$
theories and up to rank 4 for  $\CN=8$ except $F_4$. This gives us
enough confidence to believe the heuristic arguments presented
above hold true. Here we display the resulting $\Omega$'s explicitly.

\subsubsection*{$\CN=4$}

For $SU(N)$ with the permutation group $S_N$ as the Weyl group,
the only elliptic Weyl elements are the fully
cyclic ones, such as $(123\cdots N)$ that permutes
$1\rightarrow 2\rightarrow 3\rightarrow\cdots\rightarrow N\rightarrow 1.$ There are $(N-1)!$ number
of such an $N$-cyclic permutations, belonging to a single
conjugacy class, each of which contributes
$$\frac{1}{{\rm Det}\left({\bf y}^{-1}-{\bf y}\cdot w\right)}=\frac{1}{ {\bf y}^{-N+1}+{\bf y}^{-N+3}+\cdots+ {\bf y}^{N-1}} \ ,$$
where one evaluated the left hand side on the irreducible
$(N-1)$ dimensional representation. This gives
\begin{itemize}
\item $SU(N)$
\bea
\Omega_{\cN=4}^{SU(N)}({\bf y})=
\frac{1}{|W|}\sum'_{w}\frac{1}{{\rm Det}\left({\bf y}^{-1}-{\bf y}\cdot w\right)}
=\frac{1}{N}\frac{{\bf y}^{-1}-{\bf y}}{{\bf y}^{-N}-{\bf y}^N} \ .
\eea
\end{itemize}
For more general gauge groups, however, there will
be more than one such conjugacy classes contributing.

For $SO(2N+1)$ and $Sp(N)$ and for $SO(2N)$, the Weyl
groups are $S_Z\ltimes (Z_2)^N$ and $S_N\ltimes (Z_2)^{N-1}$,
respectively. The elliptic Weyl elements have the general form,
$$(\dot a \dot b \dot c d\dots)(kl\dot m n\dots)\cdots $$
where $(\dots)$ again represents a cyclic permutation and each dot
above a label means a sign flip. The number of such dots in each cyclic factor
must be all odd. For $SO(2N)$, the total number of
dots is even. The relevant determinant factorizes in terms of
those of such cyclic factors,
$$
\left.\frac{1}{{\rm Det}\left(y^{-1}-y\cdot w\right)}\right\vert_{w=(\dot 1,\dot 2,\cdots, \dot k, k+1,\cdots,n)}
= \frac{1}{{\bf y}^{-n}+(-1)^{k+1}{\bf y}^n}\ ,
$$
which also shows why we need odd $k$ for $w$ to be elliptic.
Once the $n$ labels are determined and $k$ fixed, we can
count elements within such a class. Permutation among
the chosen $n$ labels gives $(n-1)!$ possible maximally
cyclic permutation among these $n$ labels, and for each
choice, assignment of dots gives a further $_nC_k$ choices.
Unlike $SU(N)$, there seems to be no simple and closed
form for general ranks, so here we list the answer for
ranks up to five.

\begin{itemize}

\item  $SO(4)$
\bea
\Omega_{\cN=4}^{SO(4)}({\bf y})=\frac14 \frac{1}{({\bf y}^{-1}+{\bf y})^2}
\eea
\item $SO(5)$ and $Sp(2)$
\bea
\Omega_{\cN=4}^{SO(4)}({\bf y})=\Omega_{\cN=4}^{Sp(2)}({\bf y})=
\frac{1}{8}\left[\frac{2}{{\bf y}^{-2}+{\bf y}^2}+\frac{1}{({\bf y}^{-1}+{\bf y})^2}\right]
\eea
\item $SO(6)$
\bea
\Omega_{\cN=4}^{SO(6)}({\bf y})=\frac{1}{24}\frac{6}{({\bf y}^{-2}+{\bf y}^2)({\bf y}^{-1}+{\bf y})}
\eea
\item $SO(7)$ and $Sp(3)$
\begin{eqnarray}
&&\Omega_{\cN=4}^{SO(7)}({\bf y})=\Omega_{\cN=4}^{Sp(3)}({\bf y})=\cr\cr
&&\frac{1}{48}\left[\frac{8}{{\bf y}^{-3}+{\bf y}^3}+ \frac{ 6 }{({\bf y}^{-2}+{\bf y}^2)({\bf y}^{-1}+{\bf y})}
+\frac{1}{({\bf y}^{-1}+{\bf y})^3}\right]
\end{eqnarray}
\item $SO(8)$
\begin{eqnarray}
&&\Omega_{\cN=4}^{SO(8)}({\bf y})=\cr\cr
&&\frac{1}{192}\left[ \frac{32}{({\bf y}^{-3}+{\bf y}^3)({\bf y}^{-1}+{\bf y})}
+\frac{ 12 }{({\bf y}^{-2}+{\bf y}^2)^2}+
\frac{1}{({\bf y}^{-1}+{\bf y})^4}\right]
\end{eqnarray}
\item $SO(9)$ and $Sp(4)$
\begin{eqnarray}
&&\Omega_{\cN=4}^{SO(9)}({\bf y})=\Omega_{\cN=4}^{Sp(4)}({\bf y})=\cr\cr
&&\frac{1}{384}\left[\frac{48}{{\bf y}^{-4}+{\bf y}^4}+
 \frac{32}{({\bf y}^{-3}+{\bf y}^3)({\bf y}^{-1}+{\bf y})}\right.\cr\cr
&&\hskip 1.5cm\left. + \frac{12}{({\bf y}^{-2}+{\bf y}^2)^2} +\frac{12}{({\bf y}^{-2}+{\bf y}^2)({\bf y}^{-1}+{\bf y})^2}
+\frac{1}{({\bf y}^{-1}+{\bf y})^4}\right]
\end{eqnarray}
\item $SO(10)$
\begin{eqnarray}
&&\Omega_{\cN=4}^{SO(10)}({\bf y})=\cr\cr
&&\frac{1}{1920}\left[ \frac{240}{({\bf y}^{-4}+{\bf y}^4)({\bf y}^{-1}+{\bf y})}\right.\cr\cr
&&\hskip 1.5cm\left. +  \frac{160}{({\bf y}^{-3}+{\bf y}^3)({\bf y}^{-2}+{\bf y}^2)}  + \frac{20 }{({\bf y}^{-2}+{\bf y}^2)({\bf y}^{-1}+{\bf y})^3}
\right]
\end{eqnarray}
\item $SO(11)$ and $Sp(5)$
\begin{eqnarray}
&&\Omega_{\cN=4}^{SO(11)}({\bf y})=\Omega_{\cN=4}^{Sp(5)}({\bf y})=\cr\cr
&&\frac{1}{3840}\left[\frac{384}{{\bf y}^{-5}+{\bf y}^5}+
 \frac{240}{({\bf y}^{-4}+{\bf y}^4)({\bf y}^{-1}+{\bf y})}+  \frac{160}{({\bf y}^{-3}+{\bf y}^3)({\bf y}^{-2}+{\bf y}^2)} \right.\cr\cr
&&\hskip 1.5cm
+\frac{80}{({\bf y}^{-3}+{\bf y}^3) ({\bf y}^{-1}+{\bf y})^2} +\frac{60}{({\bf y}^{-2}+{\bf y}^2)^2({\bf y}^{-1}+{\bf y})}
\cr\cr
&&\hskip 1.5cm
\left.
+\frac{20}{({\bf y}^{-2}+{\bf y}^2)({\bf y}^{-1}+{\bf y})^3} +\frac{1}{({\bf y}^{-1}+{\bf y})^5}\right]
\end{eqnarray}

\end{itemize}
For exceptional groups, the Weyl groups and $\cI^{\CO(G)}_{\rm bulk}$ are
more complicated. Here we display the results for $G_2$ and $F_4$, whose Weyl
group and elliptic Weyl elements are also summarized in Appendix~\ref{C}.
\begin{itemize}
\item $G_2$
\begin{eqnarray}
\Omega_{\cN=4}^{G_2}({\bf y})=\frac{1}{12}\left[ \frac{2}{{\bf y}^{-2}-1+{\bf y}^2}
+  \frac{2}{{\bf y}^{-2}+1+{\bf y}^2}  + \frac{1}{({\bf y}^{-1}+{\bf y})^2}
\right]
\end{eqnarray}
\item $F_4$
\begin{eqnarray}
&&\Omega_{\cN=4}^{F_4}({\bf y})=\cr\cr
&&\frac{1}{1152}\left[
\frac{144}{{\bf y}^{-4} + {\bf y}^4}
+ \frac{96}{{\bf y}^{-4} - 1 + {\bf y}^4}
+ \frac{64}{{\bf y}^{-4} + {\bf y}^{-2}+{\bf y}^2 + {\bf y}^4} \right.\cr\cr
&&\hskip 1cm + \frac{12}{({\bf y}^{-2} + {\bf y}^2)^2}
+ \frac{16}{({\bf y}^{-2} - 1 + {\bf y}^2)^2}
+ \frac{16}{({\bf y}^{-2} + 1 + {\bf y}^2)^2}\cr\cr
&&\hskip 1cm
 \left.+ \frac{36}{({\bf y}^{-1} + {\bf y})^2 ({\bf y}^{-2} + {\bf y}^2)}
+ \frac{1}{({\bf y}^{-1} + {\bf y})^4}
\right]
\end{eqnarray}

\end{itemize}

\subsubsection*{$\CN=8$}

Their analogues for $\CN=8$ theories are equally straightforward.
Given $\Omega^{G}_{\cN=4}$ expressed as a sum  over the conjugacy
classes, $\cC$, of elliptic Weyl element in $W(G)$,
$$\Omega^{G}_{\cN=4}=\sum_{\cC}'\frac{A_\cC}{P_\cC({\bf y})} \ , $$
with rational numbers $A_\cC$ and the symmetric Laurent polynomials
$P_\cC$, of degree $r$,
its $\cN=8$  counterpart is given as
\bea\label{N=8'}
\quad\Omega^{G}_{\cN=8}=\sum_{\cC}'\frac{A_\cC}{P_\cC({\bf y})}
\cdot\frac{P_\cC(x^{-1/2}{\bf y})}{P_\cC(x^{-1/2})} \ ,
\eea
with the unit flavor charge and zero $R$-charge for the adjoint chiral.
For instance, $SO(4)$ case works as
\bea
\Omega^{SO(4)}_{\cN=4}=\frac{1/4}{({\bf y}^{-1}+{\bf y})^2}
\quad\rightarrow\quad
\Omega^{SO(4)}_{\cN=8}=\frac{1/4}{({\bf y}^{-1}+{\bf y})^2}
\cdot\frac{(x^{1/2}{\bf y}^{-1}+x^{-1/2}{\bf y})^2}{(x^{1/2}+x^{-1/2})^2} \ .
\eea

\subsection{${\cal N}=16$ Pure Yang-Mills }\label{s4.2}

With the maximal supersymmetry of $\CN=16$, threshold bound states
are in general expected, as $SU(N)$ case corresponds to the well-known
multi-D0 bound state problem~\cite{Witten:1995ex}. Therefore the twisted
partition function must include integral contribution from such $L^2$
states. This also means that the continuum sectors can be more involved
for higher rank theories, since more than one partial bound states may
form and new continuum contributions from such sectors can enter $\Omega=\cI_{\rm bulk}$.

For $SU(N)$, at the numerical level, $\cI_{\rm bulk}$ and its general
structure has been identified long time ago,
which motivated this study to begin with. The numerical limit
of the twisted partition function was known to be~\cite{Moore:1998et}
\begin{equation}\label{16'}
\Omega^{SU(N)}_{\CN=16}\biggr\vert_{{\bf y} \rightarrow 1;x\rightarrow 1}
=\sum_{p\vert N}\frac{1}{p^2} \ .
\end{equation}
We have rederived this expression by explicit computation
of fully equivariant version for some low-rank examples.
With three adjoint chirals $\Phi_a$'s, we can assign
$R$-charges $(R_1,R_2,R_3)$ and turn on one U(1) flavor chemical
potential with charges $(F_1,F_2,F_3)$, so that the leading
superpotential is trilinear $\sim \Phi_1\Phi_2\Phi_3$, and find
\begin{eqnarray}\label{16equivariant'}
\Omega^{SU(N)}_{\CN=16}
&=&1+ \sum_{p\vert N;p>1}\Delta^{SU(p)}_{\CN=16}\ ,
\end{eqnarray}
where $\Delta^G_{\CN=16}$ are the direct analog of the expressions for
$\Omega^G_{\CN=8}$ on the right hand side of eq.~(\ref{N=8'}).
They are sums over conjugacy classes, $\cC$, of elliptic Weyl
elements of $G$,
\bea\label{Delta16}
\Delta^{G}_{\cN=16}\equiv\sum_{\cC}'\frac{A_\cC}{P_\cC({\bf y})}\cdot
\prod_{a=1}^3 \frac{P_\cC(x^{F_a/2}{\bf y}^{(R_a-2)/2})}{P_\cC(x^{F_a/2}{\bf y}^{R_a/2})}\ .
\eea
Note that the limit $x\rightarrow 1$ is well-defined.
This happy outcome, however, may be attributed to the fact that
we are cheating a little here for $\CN=16$. The
trilinear superpotential $\sim \Phi_1\Phi_2\Phi_3$
does not guarantee the commutator superpotential ${\rm tr}\, (\Phi_1[\Phi_2,\Phi_3])$.
Thus, in principle, we are computing $\Omega$'s for a deformed
version of $\CN=16$ pure Yang-Mills theory. Nevertheless, we believe
this suffices for the purpose of Witten Index computation.

For one thing, the physical origin of eq.~(\ref{16'}), now
elevated to its equivariant version~(\ref{16equivariant'}), has been understood
in very physical terms: $p=1$ gives $1$, corresponding to the
genuine $L^2$ index of this quantum mechanics,
while the others with $p>1$ come from asymptotic dynamics
of a sector where the $p$ partial bound states associated
with $p$ number of $SU(N/p)$ subsystems are separated far
from one another. In other words, $\Omega^{SU(N)}_{\CN=16}$
has the simple interpretation in terms of Witten indices as
\begin{equation}\label{16}
\Omega^{SU(N)}_{\CN=16}\biggr\vert_{{\bf y} \rightarrow 1;x\rightarrow 1}\;=\;\cI_{\CN=16}^{SU(N)}\biggr\vert_{{\bf y} \rightarrow 1}+
\sum_{p\vert N; p>1}\frac{1}{p^2}\times \cI_{\CN=16}^{SU(N/p)}\biggr\vert_{{\bf y} \rightarrow 1} \ ,
\end{equation}
with $\cI^{SU(N)}_{\CN=16}=1$ for all $N$. The insight
that anticipated and explained this formula,
originates in Ref.~\cite{Yi:1997eg}.
This intriguing structure of the twisted partition function,
or the bulk part of the Witten index, persists well
beyond pure Yang-Mills theories, as will be explained
in the next section. Here it suffices to note that, once we understand
these structures of fractional $\Omega$'s, true and integral
$L^2$ index can be read off easily from $\Omega=\cI_{\rm bulk}$,
without having to deal with the very subtle boundary contribution $\delta\cI$.

We repeated the same exercise for other simple gauge groups also.
The fully equivariant $\Omega^G_{\CN=16}$'s are in Eq.~(\ref{otherN=16})
below, while the table lists the numerical limits for $\CN=4,8,16$ for easy
comparisons among various theories. See the table below.
In the past, there have been attempts
to compute these numbers. After $\CN=4,8,16$ $SU(2)$ cases were
computed \cite{Yi:1997eg,Sethi:1997pa}, Moore, Nekrasov,
and Shatashvili (MNS) gave a localization prescription
for $SU(N)$~\cite{Moore:1998et}, with the same result as above.
However, later adaptations of MNS for other gauge groups
\cite{Staudacher:2000gx, Pestun:2002rr} do not agree with the
above. Our computation here should be taken to supersede these
older results; Note that the localization procedure
of HKY {\it derives} the zero mode contour while MNS gave
an intelligent guess on the contour. Failure of the various
contour ``prescriptions" offered in the past is hardly surprising,
especially for theories with degenerate singularities for which
order of integration is crucial and mutually distinct between
different degenerate singularities. Such subtleties manifest
also among $U(N)$ based theories, with the simplest example
we know of being non-Abelian cyclic triangle quivers\cite{Chiung}.

\vskip 5mm
\begin{equation}\nonumber
\begin{array}{c|cc}
& \qquad{\cal N}=4,8\qquad &\qquad{\cal N}=16 \qquad\\ \\ \hline \\
\qquad SU(N) \qquad & \frac{1}{N^2} &\sum_{p\vert N}\frac{1}{p^2} \\ \\ \hline \\
SO(4) & \frac{1}{16} & \frac{25}{16}  \\ \\
SO(6)=SU(4) & \frac{1}{16} & \frac{21}{16} \\ \\
SO(8) & \frac{59}{1024} & \frac{3755}{1024}\\ \\\hline\\
SO(5) & \frac{5}{32} & \frac{53}{32} \\ \\
SO(7) & \frac{15}{128} & \frac{267}{128} \\ \\
SO(9) & \frac{195}{2048} & \frac{7555}{2048} \\ \\ \hline \\
Sp(2) & \frac{5}{32} & \frac{53}{32} \\ \\
Sp(3) & \frac{15}{128} & \frac{395}{128} \\ \\
Sp(4) & \frac{195}{2048} & \frac{8067}{2048} \\ \\ \hline \\
G_2 & \frac{35}{144} &  \frac{395}{144} \\ \\
\end{array}
\end{equation}
\vskip 5mm\noindent

The analogues of eq.~(\ref{16equivariant'}), complete with
how $\Omega$ is decomposed in terms of continuum sectors
associated with subgroups, are more involved for general
gauge groups. For the above low-rank examples,
the following decompositions are found,

\bea\label{otherN=16}
\Omega^{SO(5)/Sp(2)}_{\CN=16}  &=& 1 + 2\Delta^{SO(3)/Sp(1)}_{\CN=16}
+ \Delta^{SO(5)/Sp(2)}_{\CN=16} \ ,\cr\cr
\Omega^{G_2}_{\CN=16} &=& 2 + 2\Delta^{SU(2)}_{\CN=16}
+ \Delta^{G_2}_{\CN=16} \ ,\cr\cr
\Omega^{SO(7)}_{\CN=16} &=& 1 + 3\Delta^{SO(3)}_{\CN=16}
+  \left(\Delta^{SO(3)}_{\CN=16}\right)^2 + \Delta^{SO(5)}_{\CN=16} + \Delta^{SO(7)}_{\CN=16} \ ,\cr\cr
\Omega^{Sp(3)}_{\CN=16} &=& 2 + 3\Delta^{Sp(1)}_{\CN=16}
+ \left(\Delta^{Sp(1)}_{\CN=16}\right)^2 + \Delta^{Sp(2)}_{\CN=16} + \Delta^{Sp(3)}_{\CN=16} \ ,\cr\cr
\Omega^{SO(8)}_{\CN=16} &=& 2 + 4\Delta^{SO(3)}_{\CN=16} + 2\left(\Delta^{SO(3)}_{\CN=16}\right)^2 +
\left(\Delta^{SO(3)}_{\CN=16}\right)^3  + 3\Delta^{SO(5)}_{\CN=16} + \Delta^{SO(8)}_{\CN=16} \ ,\cr\cr
\Omega^{SO(9)}_{\CN=16} &=& 2 + 4\Delta^{SO(3)}_{\CN=16} + 2\left(\Delta^{SO(3)}_{\CN=16}\right)^2 +
2\Delta^{SO(5)}_{\CN=16}+ \Delta^{SO(3)}_{\CN=16}\cdot\Delta^{SO(5)}_{\CN=16} +\Delta^{SO(7)}_{\CN=16}
 +\Delta^{SO(9)}_{\CN=16} \ , \cr\cr
\Omega^{Sp(4)}_{\CN=16} &=& 2 + 5\Delta^{Sp(1)}_{\CN=16} + 2\left(\Delta^{Sp(1)}_{\CN=16}\right)^2 +
2\Delta^{Sp(2)}_{\CN=16}+ \Delta^{Sp(1)}_{\CN=16}\cdot\Delta^{Sp(2)}_{\CN=16} +\Delta^{Sp(3)}_{\CN=16}
 +\Delta^{Sp(4)}_{\CN=16} \ ,\nonumber \\
&&
\eea
with $\Delta^G_{\CN=16}$ defined in eq.~(\ref{Delta16}).\footnote{Recall that $\Delta^G$ depends only
on the Weyl group and its action on the Cartan subalgebra of $G$,
allowing the identities, $\Delta^{SO(2N+1)}=\Delta^{Sp(N)}$, as well as
more trivial ones following from the equality of the Lie Algebras,
$\Delta^{SO(3)}=\Delta^{SU(2)}=\Delta^{Sp(1)}$, $\Delta^{SO(5)}=\Delta^{Sp(2)}$,
$\Delta^{SO(4)}=(\Delta^{SU(2)})^2$, and $\Delta^{SO(6)}=\Delta^{SU(4)}$.}
Just as in $SU(N)$ cases, where each fractional term involving
$\Delta^{SU(p)\subset SU(N)}_{\CN=16}$'s is from a specific continuum sector,
we can again blame each term with $\Delta^{\tilde G\subset G}_{\CN=16}$'s
to a specific continuum sector with partial bound states,
and thus read off the bona-fide Witten index as

\bea
\cI^{SO(5)=Sp(2)}_{\CN=16} &=& 1 \ , \cr\cr
\cI^{G_2}_{\CN=16} &=& 2 \ ,    \cr\cr
\cI^{SO(7)}_{\CN=16} &=& 1 \ , \cr\cr
\cI^{Sp(3)}_{\CN=16} &=& 2 \ , \cr\cr
\cI^{Sp(8)}_{\CN=16} &=& 2 \ , \cr\cr
\cI^{SO(9)}_{\CN=16} &=& 2 \ , \cr\cr
\cI^{Sp(4)}_{\CN=16} &=& 2 \ ,
\eea
which again count $L^2$ threshold bound states.

Interestingly, the same numbers of states have been
advocated in the past by Kac and Smilga \cite{Kac:1999av},
who inferred these numbers by studying mass-deformed
versions of such quantum mechanics \cite{Porrati:1997ej}.
Of course this approach
of theirs, where the ground states are classified as
classical solutions, is yet another, rather heavy-handed,
attempt to regulate the infrared subtleties; the result
cannot be considered conclusive on its own. Nevertheless, our Witten
index computation, with very different and much milder infrared
regulator and in particular with the ability to keep track of
various continuum contributions to $\cI^{G}_{\rm bulk}$ sector
by sector, confirms their counting for these low rank examples.

\section{Index and Rational Invariant for $\CN=4$  Quiver Theories}\label{quiver}

Let us start by recalling the pure $SU(N)$ Yang-Mills quantum
mechanics with $\cN=16$ supercharges, where the localization procedure gives
\bea
\Omega_{{\cal N}=16}^{SU(N)}\biggr\vert_{{\bf y} \rightarrow 1;x\rightarrow 1}
=\sum_{p\vert N}\frac{1}{p^2} \ .
\eea
As we already noted, this fractional result for the bulk part of
Witten index has a well-known interpretation as in eq.~(\ref{16}),
implying
\bea
\cI_{{\cal N}=16}^{SU(N)}\;\;=\;\;\cI_{{\cal N}=16}^{SU(N)}\biggr\vert_{{\bf y} \rightarrow 1}=1 \ ,
\eea
for all $N$.
In fact, the same is true of the somewhat more trivial cases with $\CN=4,8$,
so we really have
\bea
\Omega_{{\cal N}=4,8,16}^{SU(N)}\biggr\vert_{{\bf y} \rightarrow 1;x\rightarrow 1}
\;=\;\cI_{\CN=4,8,16}^{SU(N)}\biggr\vert_{{\bf y} \rightarrow 1}+
\sum_{p\vert N; \,p>1}\frac{1}{p^2}\times \cI_{\CN=4,8,16}^{SU(N/p)}\biggr\vert_{{\bf y} \rightarrow 1} \ .
\eea
Comparing with the actual expression of $\Omega$'s computed by the
localization tells us immediately that
\bea
\cI_{\CN=4,8}^{SU(N)}\;\;=\;\;\cI_{\CN=4,8}^{SU(N)}\biggr\vert_{{\bf y} \rightarrow 1}&=&0\ ,\cr\cr
\cI_{\CN=16}^{SU(N)}\;\;=\;\; \cI_{\CN=16}^{SU(N)}\biggr\vert_{{\bf y} \rightarrow 1}&=&1 \ .
\eea
For $\CN=4,8$, the only surviving term is $p=N$,
\bea
\Omega_{{\cal N}=4,8}^{SU(N)}\biggr\vert_{{\bf y} \rightarrow 1;x\rightarrow 1}
\;=\;\frac{1}{N^2}\times \cI_{\CN=4,8}^{SU(1)}\biggr\vert_{{\bf y} \rightarrow 1}\;=\;\frac{1}{N^2}
\eea
where the formal statement, $\cI_{N=4,8,16}^{SU(1)}=1$, refers to
the fact the elementary vector multiplet is the basic BPS multiplet
for each case. For $\CN=4,8$, a similar line of thought leads us to
\bea\label{index48}
\cI_{\CN=4,8}^G=0 \ ,
\eea
for all simple gauge groups: This was in fact one of the basic
assumptions that lead us to the formulae (\ref{N=4}) and (\ref{N=8}),
to begin with, so our confirmation of the latter by an explicit
localization computation shows (\ref{index48}) self-consistently.

Such relations between $\cI$ and $\Omega$ are not confined
to these pure Yang-Mills quantum mechanics. While we do not
understand them generally, there is at least one very general
class of theories where such a pattern is quite prevalent, namely,
$\CN=4$ quiver quantum mechanics. For this, we first recall the so-called
rational invariant for a BPS state of fixed charge $\Gamma$,
\bea
\Omega^\Gamma({\bf y}) =\sum_{p\vert \Gamma}
\frac{{\bf y}-{\bf y}^{-1}}{p\cdot ({\bf y}^p-{\bf y}^{-p})}\times \cI^{\Gamma/p}({\bf y}^p) \ ,
\eea
where $p$ is a positive integer such that $\Gamma/p$ is
a properly quantized charge (an integral rank vector). This
 has a known inversion formula,
\bea
\cI^\Gamma({\bf y}) =\sum_{p\vert \Gamma}
\frac{ \mu(p)\cdot({\bf y}-{\bf y}^{-1})}{p\cdot ({\bf y}^p-{\bf y}^{-p})}\times
\Omega^{\Gamma/p}({\bf y}^p)\ ,
\eea
where $\mu(p)$ is the same M\"obius function that appeared briefly
in Section~\ref{s3.4} in an entirely different context.

Note that this is precisely the type of relations we saw for $SU(N)$
pure Yang-Mills quantum mechanics with various numbers of supersymmetries.
Interestingly, these rational invariants have also surfaced more
recently in the context of various wall-crossing formulae, say,
either those derived from low energy dynamics of Seiberg-Witten dyons
~\cite{Manschot:2010qz, Manschot:2011xc,Lee:2011ph,Kim:2011sc}
or those based on the mathematics of Donaldson-Thomas invariants and
Kontsevich-Soibelman wall-crossing algebra~\cite{KS, JS, GMN1, GMN2}.
Below we will review how the index of a quiver theory is organized
in terms of such rational invariants, and how this leads directly to
$L^2$ Witten index that counts threshold bound states.

\subsection{Rational Invariants and Primitive Quivers }

Recall the quantum mechanical GLSM of quiver type with
$\cN=4$ supersymmetries. A quiver diagram $\cQ$ is a pair $(V, E)$,
where $V$ is a finite set of vertices (or nodes) and $E$ is a finite set of
oriented edges connecting those vertices. We associate a physical
theory to $\cQ$ as follows. Firstly, each vertex $v_i \in V$,
$i=1, \cdots, n$, is labelled by a rank $N_i$, representing a $U(N_i)$
vector multiplet, and is equipped with an FI constant $\zeta_i$
for its trace $U(1)$. Secondly, each oriented edge $e \in E$
represents a chiral multiplet in the bifundamental representation,
$(\bar N_{t_e}, N_{h_e})$, where $t_e$ and $h_e$ are
the tail and the head vertices of the edge $e$, respectively.
One may describe $E$ via the adjacency matrix, $b=\left[b_{ij}\right]$,
which counts the arrows from node $v_i$ to $v_j$. We will consider
quivers without a flavor node, so the total gauge group is $[\prod U(N_i)]/U(1)$
after the free $U(1)$ is factored out.

In particular, we are confining our attention to
those quivers without 1-loop and 2-loop, meaning, respectively, an
edge that starts and ends at the same node and edges between a pair
of nodes with mutually
opposite orientations. The former, $b_{ii}=0$, means that there is no adjoint
chirals, and the latter, $b_{ij}\cdot b_{ji}=0$ for $i\neq j$, means
any pair of chirals with mutually opposite gauge charges are taken to
annihilate each other. This class of quivers arises naturally as low
energy dynamics of BPS objects in $d=4$ $\CN=2$ theories~\cite{Denef},
and has been studied quite extensively in recent years.\footnote{For
quivers without a loop, the problem has been worked out  extensively and
repeatedly. A pioneering work by Reineke \cite{Reineke} gave a closed-form
formula for the Poincare polynomial of a Higgs vacuum moduli space,
while a series of study by Manschot, Pioline, and Sen offered
a very comprehensive fixed-point proposal via the Coulombic multi-center
picture \cite{Manschot:2010qz,Manschot:2010qz}. Then, Ref.~\cite{Kim:2011sc}
re-derived these formulae from physical low energy dynamics of
$d=4$ $\CN=2$ dyons and also much clarified the end result mathematically.
The results of the aforementioned analyses, as well as that of the wall-crossing
formula due to Kontsevich and Soibelman, were subsequently shown to
be equivalent among one another \cite{Sen:2011aa}.

However, for quivers with a loop, the story turned out to be far more
involved as one must consider an entirely new classes of BPS states
\cite{Bena:2012hf,Lee:2012sc} that cannot be captured by the multi-center
construction of BPS states.  Geometric and physical meaning of such
states, called ``pure Higgs" or ``intrinsics Higgs," were clarified
in Ref.~\cite{Lee:2012naa} and subsequently incorporated into the
multi-center counting in Refs.~\cite{Manschot:2012rx,Manschot:2013sya,Manschot:2014fua}.
For quivers, the index associated with such states is called the
quiver-invariant to distinguish it from Witten index. The latter
changes discontinuously across marginal stability walls, while
the former does not.

An orthogonal approach to the BPS spectra problem, very useful
for low rank $d=4$ $\CN=2$ field theories, has also been used
fruitfully, whose modern form can be traced to Ref.~\cite{GMN2}.
A representative of such works can be found in Refs.~\cite{Alim,Gaiotto:2012rg}.}

For the simplest class of quivers where the rational invariant enters,
consider primitive quivers $\cQ$ endowed with the coprime ranks $N_i$ as well as
the FI constants $\zeta_i$, and {\it without any oriented loops}. The primitivity
means $\{N_i\}$ has no nontrivial common divisor. As we will discuss later,
this setting can actually be extended to general quivers with the understanding that
the generalization to nonprimitive case gives the ``bulk" part of the index.
An Abelianization formula has been proposed \cite{Manschot:2010qz} and proved \cite{Kim:2011sc}
for this class of theories, which can be summarized compactly as follows~\cite{Lee:2013yka},
\beq
\Omega_\cQ^\zeta = \sum_{\cP} \frac{1}{|\Gamma_\cP|}
\cdot \Omega_{\cQ_\cP}^{\zeta_\cP}\cdot \Omega_\cP  \ .
\eeq
The summation here is over the partitions $\cP$ of the rank vector
$(N_1, \cdots, N_n)$, specifying a partition of each rank $N_i$ into $l_i$ positive integers,
\beq\label{partition}
\cP = (\{N_{1, a_1}\}_{a_1=1}^{l_1} \ , \cdots, \{N_{n, a_n}\}_{a_{n}=1}^{l_{n}})
\ , \quad {\text{with}}~N_i = \sum_{a_i = 1}^{l_i} N_{i,a_i} \ .
\eeq
Then, to each such partition $\cP$ is associated an Abelian quiver, $\cQ_\cP$,
with $U(1)$ vertices $v_{i,a_i}$ for
$a_i=1, \cdots, l_i,~i=1, \cdots, n$ and FI constants
$\zeta_{i,a_i}=N_{i,a_i} \zeta_i$, collectively denoted as $\zeta_\cP$.
The edge set of $\cQ_\cP$ is specified by its adjacency, so that there
are $N_{i, a_i} N_{j, a_j} b_{ij}$ edges from the node $v_{i,a_i}$ to
$v_{j,a_j}$. Simply put, this induced Abelian quiver $\cQ_\cP$
results if we partition $N_i=\sum_{a_i}N_{i,a_i}$ and treat each
$N_{i,a_i}\times N_{i,a_i}$ block as if it is a single Abelian node.
Now, the discrete group $\Gamma_\cP$ is defined by its trivial
permutation action on the nodes of $\cQ_\cP$ in the following
sense: If in the partition $N_i=\sum_{a_i} N_{i,a_i}$ we
find $N_{i,s_i}$ repeated $d_{s_i}$ times, such that
$N_i=\sum_{s_i} d_{s_i}N_{i,s_i}$ with $N_{i,s_i}\neq N_{i,s'_i}$
for $s_i\neq s'_i$, we have\footnote{For example,
with a two-node quiver with the rank vector
$(N_1,N_2)=(16,9)$ and the partition $\cP: (16,9)\rightarrow(2+2+2+5+5, \;1+1+1+1+1+4)$,
the corresponding discrete group is give as
$$\Gamma_\cP=S(3)\times S(2)\times S(5)\times S(1)\ . $$}
\bea
\Gamma_\cP= \prod_i\prod_{s_i} S(d_{s_i})\ .
\eea
The two other factors in the summand are themselves twisted
partition functions on their own.
The former factor, $\Omega_{\cQ_\cP}^{\zeta_\cP}$, is the twisted partition
function of the Abelian theory associated to
$\cQ_\cP$ with FI constants $\zeta_\cP$.
On the other hand, the latter factor is a product of many twisted partition functions
\beq
\Omega_\cP = \prod_{i=1}^n \prod_{a_i=1}^{l_i}
\frac{1}{N_{i,a_i}}\frac{{\bf y}-{\bf y}^{-1}}{{\bf y}^{N_{i,a_i}}-{\bf y}^{-N_{i,a_i}}} \ ,
\eeq
which we recognize as
\beq
\Omega_\cP = \prod_{i=1}^n \prod_{a_i=1}^{l_i}
\Omega^{SU(N_{i,a_i})}_{\CN=4} \ ,
\eeq
from the previous section.

In the Coulombic picture of the wavefunctions, which is known to be
effective for such loopless quivers, each summand has a clear physical
interpretation as follows: A sector labeled by $\cP$ corresponds to
the partial symmetry breaking,
\bea U(N_i)\rightarrow U(1)^{l_i}\times \prod_{a_i} SU(N_{i,a_i}) \ , \eea
where the low energy dynamics are then locally a product of $\cQ_\cP$
with the gauge group, $\left[\prod_{i} U(1)^{l_i}\right]/U(1)$,
and of many
$\CN=4$ pure Yang-Mills theory with the gauge groups $SU(N_{i,a_i})$.
The twisted partition function of the former give $\Omega_{\cQ_\cP}^{\zeta_\cP}$, while
the latter gives a product of pure $SU(N_{i,a_i})$ twisted partition
functions which act like the intrinsic degeneracy attached to each
node of $\cQ_\cP$. The discrete division by $\Gamma_\cP$ is a remnant
of the original Weyl group $\prod S(N_i)$ after the divisions due to
$S(N_{i,a_i})$ subgroups are used up in the computation of
$\Omega^{SU(N_{i,a_i})}_{\CN=4}$'s.

For primitive and connected quivers, the low energy dynamics is compact
and the  twisted partition function $\Omega$ has to be itself the Witten
index and thus integral, despite such a complicated Abelianization formula
with  fractional contributions term by term. In other words,
\bea
\cI_\cQ= \Omega_\cQ \ .
\eea
The rational invariant enters the story in the computational
middle steps via
\bea
\Omega^{SU(N_{i,a_i})}_{\CN=4}= \sum_{p\vert N_{i,a_i} }
\frac{{\bf y}-{\bf y}^{-1}}{p\cdot ({\bf y}^p-{\bf y}^{-p})}\times \cI^{SU(N_{i,a_i}/p)}_{\CN=4}=
\frac{1}{N_{i,a_i}}\frac{{\bf y}-{\bf y}^{-1}}{{\bf y}^{N_{i,a_i}}-{\bf y}^{-N_{i,a_i}}} \ .
\eea
For nonprimitive quivers, on the other hand, the rational invariant
becomes relevant for the final form of $\Omega_\cQ$ as well, since
the theory then comes with gapless asymptotic directions, much similar
to the pure Yang-Mills theories. We now turn to this case.

\subsection{Rational Invariants and Threshold Bound States}

A nonprimitive quiver is a quiver whose rank vector $\vec N$
is divisible by positive integer(s) $p\ge 2$. Such a quiver comes
with gapless asymptotic directions corresponding to separating $p$
subquivers, each denoted as $\cQ/p$, with the reduced rank vector
$\vec N/p$, from one another. They thus lead to a fractional
continuum contribution. Our experience with $\CN=4$ theories
so far suggests that the twisted partition function will produce
fractional results, but also that the results can be rephrased
in terms of integral Witten indices of these subquivers,
$\cQ/p$,
\bea
\Omega_\cQ({\bf y}) =\sum_{p\vert \cQ}
\frac{{\bf y}-{\bf y}^{-1}}{p\cdot ({\bf y}^p-{\bf y}^{-p})}\times \cI_{\cQ/p}({\bf y}^p)\ ,
\eea
or alternatively,
\bea \label{nonp}
\cI_\cQ({\bf y}) =\sum_{p\vert \cQ}
\frac{ \mu(p)\cdot({\bf y}-{\bf y}^{-1})}{p\cdot ({\bf y}^p-{\bf y}^{-p})}\times \Omega_{\cQ/p}({\bf y}^p)\ .
\eea
We propose that this relationship between the twisted partition
function $\Omega$, computed by the HKY localization method, and $\cI$,
the true Witten index, hold for general nonprimitive quivers.\footnote{In 
the context of the multi-center Coulombic index computation, where one
computes in the end certain symplectic volume via a fixed point theorem, 
the same type of relationship has been proposed. See Ref.~\cite{Manschot:2014fua}
for a compact summary. Their  approach deals 
with the problematic asymptotic direction by unphysical deformation 
of FI constants in the Abelianized middle steps, and also 
cannot compute the single center states that would be
counted by the quiver invariant \cite{Lee:2012sc,Lee:2012naa}. 
The latter are left as unknown input data \cite{Manschot:2012rx,Manschot:2013sya}, 
producing an incomplete answers when a loop is present.} 
Below, we test it explicitly for some simple quivers.

The simplest class of quivers where this can be tested is
the Kronecker quiver, with two nodes. With the rank vector
$(N,N)$, the theory has the obvious gapless asymptotic
directions along the Coulombic side. Denoting the quiver
as $\cQ^{(N,N)}_b$ where $b=b_{12}$ is the intersection number,
we should have
\bea
\cI_{\cQ^{(N,N)}_b}({\bf y}) =\sum_{p\vert N} \frac{ \mu(p)\cdot({\bf y}-{\bf y}^{-1})}{p\cdot ({\bf y}^p-{\bf y}^{-p})}
\times \Omega_{\cQ^{(N/p,N/p)}_b}({\bf y}^p)\ ,
\eea
which, of course, must be integral. Primitive $N=1$ cases reproduce
well-known indices,
\bea
\cI_{\cQ^{(1,1)}_b}({\bf y}) = \Omega_{\cQ^{(1,1)}_b}({\bf y})=(-1)^{b-1}\frac{{\bf y}^b-{\bf y}^{-b}}{{\bf y}-{\bf y}^{-1}}=(-1)^{b-1}\chi_{(b-1)/2}({\bf y}^2)\ ,
\eea
where $\chi_s(z)$ is the usual $SU(2)$ character of the spin $s$
representation. The overall sign is a consequence of the chirality
choice we made, $(-1)^{2J_3}$, and suggests that these states
are $U(1)_R$ neutral and classified via $SU(2)_R$ 
multiplets.\footnote{This feature is typical of loop-less quivers,
and is related to the so-called no exotics conjecture ~\cite{Gaiotto:2010be,DelZotto:2014bga}
for $d=4$ Seiberg-Witten theories. In the latter, our $SU(2)_R$
would be interpreted as the spatial rotation group.}
We computed $\Omega$'s and thereby deduced the physical Witten index as
$\cI$'s
\bea
\cI_{\cQ^{(2,2)}_{b=1}}^{\zeta>0}({\bf y}) &=& 0\ , \cr \cr
\cI_{\cQ^{(2,2)}_{b=2}}^{\zeta>0}({\bf y}) &=& 0\ , \cr\cr
\cI_{\cQ^{(2,2)}_{b=3}}^{\zeta>0}({\bf y}) &=& -\chi_{5/2}({\bf y}^2)\ , \cr\cr
\cI_{\cQ^{(2,2)}_{b=4}}^{\zeta>0}({\bf y}) &=& -\chi_{9/2}({\bf y}^2)-\chi_{5/2}({\bf y}^2)\ , \cr\cr
\cI_{\cQ^{(2,2)}_{b=5}}^{\zeta>0}({\bf y}) &=&-\chi_{13/2}({\bf y}^2)-2\chi_{9/2}({\bf y}^2)-\chi_{5/2}({\bf y}^2)\ ,
\eea
for $N=2$, and for $N=3$
\bea
\cI_{\cQ^{(3,3)}_{b=1}}^{\zeta>0}({\bf y}) &=& 0\ , \cr \cr
\cI_{\cQ^{(3,3)}_{b=2}}^{\zeta>0}({\bf y}) &=& 0\ , \cr \cr
\cI_{\cQ^{(3,3)}_{b=3}}^{\zeta>0}({\bf y}) &=& \chi_{5}({\bf y}^2)+\chi_{3}({\bf y}^2)\ .
\eea
States counted by these indices have to be at threshold.
They cannot be constructed by the usual Coulombic approximation
of Denef's type since this theory has a nontrivial
classical moduli space even for $\zeta<0$, and such a quantum
bound state cannot be argued away either via the usual multi-center
picture based on one-loop effective potential. Therefore, the
following fact, also confirmed using the same routine as for
$\zeta>0$ cases,
\bea
\cI_{\cQ^{(N,N)}_b}^{\zeta<0}({\bf y})=0 \ ,
\eea
is actually a nontrivial statement, confirmed by  our
explicit residue computations. Together, these results
are consistent with the general wall-crossing
algebra of Kontsevich and Soibelman.

The next simplest nonprimitive quivers are those with three nodes,
say, of ranks $(N,N,N)$. One class is $\cQ^{(N,N,N)}_{a,b}$ with
nonvanishing $b_{ij}$'s being $(b_{12}=a,b_{23}=b)$. This quiver
has only one nontrivial chamber, and we list the Witten index for
this chamber as follows
\bea
\cI_{\cQ^{(2,2,2)}_{1,1}}({\bf y}) &=& 0 \ , \cr \cr
\cI_{\cQ^{(2,2,2)}_{1,2}}({\bf y}) &=& 0 \ , \cr\cr
\cI_{\cQ^{(2,2,2)}_{1,3}}({\bf y}) &=& -\chi_{5/2}({\bf y}^2) \ , \cr\cr
\cI_{\cQ^{(2,2,2)}_{1,4}}({\bf y}) &=& -\chi_{9/2}({\bf y}^2)-\chi_{5/2}({\bf y}^2) \ , \cr\cr
\cI_{\cQ^{(2,2,2)}_{1,5}}({\bf y}) &=& -\chi_{13/2}({\bf y}^2)-2\chi_{9/2}({\bf y}^2)-\chi_{5/2}({\bf y}^2) \ , \cr\cr
\cI_{\cQ^{(2,2,2)}_{2,2}}({\bf y}) &=& -\chi_{5/2}({\bf y}^2)-\chi_{3/2}({\bf y}^2)-\chi_{1/2}({\bf y}^2) \ , \cr\cr
\cI_{\cQ^{(2,2,2)}_{2,3}}({\bf y}) &=& -\chi_{9/2}({\bf y}^2)-\chi_{7/2}({\bf y}^2)-3\chi_{5/2}({\bf y}^2)-\chi_{3/2}({\bf y}^2)-\chi_{1/2}({\bf y}^2)\ ,
\eea
for which we also found
\bea
\cI_{\cQ^{(1,1,1)}_{1,b}}({\bf y})&=&(-1)^{b-1}\chi_{(b-1)/2}({\bf y}^2)\ ,\cr\cr
\cI_{\cQ^{(1,1,1)}_{2,2}}({\bf y})&=&+\chi_{1}({\bf y}^2)+\chi_0({\bf y}^2)\ ,\cr\cr
\cI_{\cQ^{(1,1,1)}_{2,3}}({\bf y})&=&-\chi_{3/2}({\bf y}^2)-\chi_{1/2}({\bf y}^2)\ ,
\eea
along the way. Another such is $\cQ^{(N,N,N)}_{a,b,-c}$ with $(b_{12}=a,b_{23}=b,b_{13}=c)$,
where the negative sign in front of $c$ emphasizes that one of the arrows has
an opposite orientation and hence that the quiver does not have a 3-loop. In the chamber with $\zeta_1<0<\zeta_2<\zeta_3$, we computed the $\Omega$'s, from which $\cI$'s can be read off as,
\bea
\cI_{\cQ^{(2,2,2)}_{1,1,-1}}({\bf y}) &=& 0 \ , \cr\cr
\cI_{\cQ^{(2,2,2)}_{2,1,-1}}({\bf y}) &=& -\chi_{5/2}({\bf y}^2)-\chi_{3/2}({\bf y}^2)-\chi_{1/2}({\bf y}^2) \ , \cr\cr
\cI_{\cQ^{(2,2,2)}_{1,1,-2}}({\bf y}) &=& -\chi_{5/2}({\bf y}^2) \ , \cr\cr
\cI_{\cQ^{(2,2,2)}_{2,2,-1}}({\bf y}) &=& -\chi_{9/2}({\bf y}^2)-\chi_{7/2}({\bf y}^2)-3\chi_{5/2}({\bf y}^2)
-\chi_{3/2}({\bf y}^2)-\chi_{1/2}({\bf y}^2) \ , \cr\cr
\cI_{\cQ^{(2,2,2)}_{2,1,-2}}({\bf y}) &=& -\chi_{9/2}({\bf y}^2)-\chi_{7/2}({\bf y}^2)-3\chi_{5/2}({\bf y}^2)
-2\chi_{3/2}({\bf y}^2)-\chi_{1/2}({\bf y}^2)\ , \cr\cr
\cI_{\cQ^{(2,2,2)}_{3,1,-1}}({\bf y}) &=& -\chi_{9/2}({\bf y}^2)-\chi_{7/2}({\bf y}^2)-3\chi_{5/2}({\bf y}^2)
-2\chi_{3/2}({\bf y}^2)-2\chi_{1/2}({\bf y}^2)\ , \cr\cr
\cI_{\cQ^{(2,2,2)}_{1,1,-3}}({\bf y}) &=& -\chi_{9/2}({\bf y}^2)-\chi_{5/2}({\bf y}^2)\ , \cr\cr
\cI_{\cQ^{(2,2,2)}_{2,2,-2}}({\bf y}) &=& -\chi_{13/2}({\bf y}^2)-\chi_{11/2}({\bf y}^2)-4\chi_{9/2}({\bf y}^2)\cr
&&-3\chi_{7/2}({\bf y}^2)-4\chi_{5/2}({\bf y}^2)-\chi_{3/2}({\bf y}^2)-\chi_{1/2}({\bf y}^2)\ ,
\eea
and
\bea
\cI_{\cQ^{(1,1,1)}_{1,1,-1}}({\bf y}) &=& -\chi_{1/2}({\bf y}^2)\ , \cr\cr
\cI_{\cQ^{(1,1,1)}_{2,1,-1}}({\bf y}) &=& +\chi_{1}({\bf y}^2)+\chi_0({\bf y}^2) \ , \cr\cr
\cI_{\cQ^{(1,1,1)}_{1,1,-2}}({\bf y}) &=&  +\chi_{1}({\bf y}^2)\ , \cr\cr
\cI_{\cQ^{(1,1,1)}_{2,2,-1}}({\bf y}) &=&-\chi_{3/2}({\bf y}^2)-\chi_{1/2}({\bf y}^2)\ , \cr\cr
\cI_{\cQ^{(1,1,1)}_{2,1,-2}}({\bf y}) &=& -\chi_{3/2}({\bf y}^2)-\chi_{1/2}({\bf y}^2)\ , \cr\cr
\cI_{\cQ^{(1,1,1)}_{3,1,-1}}({\bf y}) &=& -\chi_{3/2}({\bf y}^2)-\chi_{1/2}({\bf y}^2)\ , \cr\cr
\cI_{\cQ^{(1,1,1)}_{1,1,-3}}({\bf y}) &=& -\chi_{3/2}({\bf y}^2)\ , \cr\cr
\cI_{\cQ^{(1,1,1)}_{2,2,-2}}({\bf y}) &=& +\chi_{2}({\bf y}^2)+\chi_{1}({\bf y}^2) \ .
\eea
Again, all of these are integral and  sums of $SU(2)$ characters, and
agree with wall-crossing formulae, further supporting our relation (\ref{nonp}).

\section{Summary}

We explored twisted partition functions $\Omega$ of $\CN\ge 4$
gauged quantum mechanics with gapped and gapless asymptotic
directions at quantum level, for the purpose of isolating Witten
index $\cI$ with proper physical boundary condition. One crucial prerequisite
for the localization procedure is the chemical potentials that
also double as infrared regulators. We have demonstrated how
the latter is often ignorant of the proper $L^2$ boundary condition,
such that the computed $\Omega$ does not easily capture the Witten
index. Of course, one should not be surprised by this at all
since the boundary conditions imposed by these two are so much
different from each other. While there exist formal arguments
how topological quantities like index should be robust, such
invariance arguments are meant to be applied to compact theories
and are well-known to fail in the presence of asymptotic
flat directions.

We also explored a more useful question of whether and how such
$\Omega$'s might contain information about the true Witten indices
$\cI$'s. For several classes of Abelian theories, with asymptotically
conical quantum moduli spaces, we have seen that different expansions
of $\Omega$ with respect to flavor chemical potentials seemingly capture
various and mutually distinct cohomologies, instead of counting proper
$L^2$ states. For these classes, one does find, {\it a posteriori}, $L^2$
ground state counting hidden as the common flavor-neutral sector under
these various expansions, for both $\CN=4$ and $\CN=8$ theories.
Non-Abelian theories offer further challenges as multi $L^2$ states
conspire to bring in extra integral contributions; a universal
prescription for extracting $\cI$ from $\Omega$ seems unlikely
for the moment.

While we explored relations between $\Omega$ and $\cI$ for noncompact
dynamics at the level of supersymmetric quantum mechanics, observations
made here are hardly confined to $d=1$. After all, the Witten index for field
theories in $d\ge 2$ dimensions often reduces to that of zero modes. Also recall that,
with four supercharges and larger, the Higgs moduli space of the
gauged dynamics is pretty much independent of the dimension. In
particular, the related issues must be confronted for elliptic genera
when the relevant superconformal field theory is not compact \cite{Harvey:2014nha}.

One additional feature for $d=1$ shows up when the asymptotically
flat direction appears along the Coulombic side. The gap along such directions
are typically controlled by the FI constant, so this type of noncompactness
is quite generic since the latter can be introduced only for $U(1)$ gauge groups.
Even for theories based on $U(N)$ gauge groups, such  flat directions open
up when some $\zeta$ approaches zero, which is responsible for the
wall-crossing behavior of Witten index \cite{HKY} for theories that are
otherwise compact. While $\Omega$'s for such noncompact theories are
often non-integral, a systematic extraction of $\cI$ from $\Omega$ is
sometimes possible via the notion of rational invariants. In physics
literature, glimpse of the latter was first seen some twenty years ago
in the context of D-brane bound state problems \cite{Yi:1997eg,Green:1997tn,Moore:1998et},
while the same notion has resurfaced more recently and more systematically
in the context of the wall-crossing problem \cite{KS,Manschot:2010qz,Kim:2011sc}.
The upshot is that, even though a gapless Coulombic direction sounds like
a worse problem than gapless Higgs directions, this is actually not the
case. Since the continuum sectors contribute to $\Omega$ in  a very
specific manner, we can easily identify and subtract  away $-\delta\cI$
sitting inside $\Omega=\cI_{\rm bulk}$ without much extra effort; no need to
compute $\delta\cI$ separately, in particular.

$\Omega=\cI_{\rm bulk}$ for such a theory is decomposed into the integral
Witten index $\cI$ and other fractional continuum contributions associated
with certain subgroups $\tilde G$ of $G$. The fractional part, $-\delta\cI$, can be
built entirely from elliptic Weyl elements of $\tilde G$'s. This map is
pretty simple for theories based on $U(N)$ or $SU(N)$ groups, such as
quiver theories, while our computation shows that a similar, more complicated,
dictionary is present for other types of gauge groups as well. In this note,
we have shown how this line of thinking leads to vanishing index for pure
Yang-Mills quantum mechanics $\cI^G_{\CN=4,8}=0$ for all simple gauge
groups $G$ and how one can extract integral $\cI^G_{\CN=16}$'s
from fractional $\Omega^G_{\CN=16}$'s. For some low-rank simple groups,
we actually isolated the integral Witten index $\cI^G_{\CN=16}$ without
a separate computation of the continuum contribution $\delta\cI$ as
a boundary term. This is a rather remarkable property of supersymmetric
gauged quantum mechanics, which may be exploited further in future.

One important and very much related matter we did not address in
this note is the quiver invariant, or more generally the GLSM invariant.
While we considered loop-less quivers in section 5, quivers with
oriented loops are known to admit the subsector of BPS spectra
that are common for all physical chambers in $\zeta$-space and
immune to wall-crossing \cite{Lee:2012sc,Lee:2012naa,Bena:2012hf}. From
other examples in Ref.~\cite{HKY}, it is also clear such wall-crossing-safe
states would be found in many GLSM's that admit superpotentials;
one obvious prototype is the GLSM that builds the familiar quintic
Calabi-Yau 3-fold in the geometric phase.

These special subset of ground states can be distinguished from
wall-crossing states as being $L^2$ normalizable even at the
walls of marginal stability.
For this reason, the GLSM (or quiver) invariant has been proposed to be
captured by the true $L^2$ Witten index of the theory sitting
exactly at $\zeta=0$ \cite{Kim:2015fba} for all FI constants,
\bea
\cI({\bf y})\biggr\vert_{\rm GLSM\;Inv}\equiv \lim_{\beta\rightarrow \infty}{\rm tr}
\left[ (-1)^{F} {\bf y}^{R+\cdots} e^{-\beta H^{\zeta=0}}\right] \ .
\eea
At such a point,
all Coulombic directions open up as gapless asymptotic directions,
similarly as in Sections 4 and 5. In fact, the $L^2$ Witten index
for $\CN=16$ gauged quantum mechanics with simple gauge groups
can be thought of as the simplest prototype of this quantity,
since the GLSM invariant would equal the Witten index for
non-wall-crossing theories. For more general theories with $\CN=4$
or less, where the GLSM invariant does not equal the Witten indices,
a better way to handle the path-integral precisely at $\zeta=0$
is needed, to which we wish to come back in near future.

\vskip 1cm
\centerline{\bf Acknowledgement}
\vskip 5mm
We are indebted to  Kentaro Hori and Heeyeon Kim for numerous discussions
and also for collaboration at an early stage of this work. PY thanks
Francesco Benini, Richard Eager, Chiung Hwang, Bumsig Kim, Sungjay Lee,
Boris Pioline, and Edward Witten
for useful conversations. SJL would like to thank Korea Institute for Advanced
Study and Northwest University for hospitality during various stages of
this work. The work of SJL is supported in part by NSF grant PHY-1417316.
\vskip 1cm
\appendix

\section{Jeffrey-Kirwan Residue}\label{JKR}
\subsection{Constructive JK Residue and Twisted Partition Function}\label{A}

The aim of this appendix is to summarize our algorithmic
methodology for the localization computation~\eqref{loc} of twisted
partition function, based on a constructive definition of the
JK residue~\cite{SV}. Here, we will give a detailed description of the procedure
to compute the internal part contribution.\footnote{The asymptotic
part can be taken care of along exactly the same line, except that
one also needs to take into account the additional, fictitious charge
$Q_{\infty}=-\zeta$ for the flag combinatorics.
In many cases, however, one may argue that such an asymptotic
part never contributes. Indeed, we could
obtain all the results in this note without ever considering the asymptotic part.}
It turns out that the procedure is systematic enough
to be implemented on a computer. Indeed, we developed
the relevant Mathematica routines and used them
extensively to obtain many of the results in this note.

To set the notations up, let us recall that the integrand $g(u)$,
defined by eqs.~\eqref{g}, \eqref{gv}, and \eqref{gm}, have singularities
along the following hyperplanes,
\bea\label{hv}
H_{{\rm vector}}^{\alpha} &=& \{\,u\,|\, \alpha \cdot u-z =0\,\} \ , \quad \alpha\in\Delta_G \ , \\ \label{hm}
H_{{\rm matter}}^{a, \rho} &=& \{\,u\,|\, \rho \cdot u + \frac{R_a}{2} z + F_a \cdot \mu = 0 \,\} \ ,
\quad\rho \in \cR_a \ , \, a=1, \cdots, A \ .
\eea
We denote the set of charges for all the physical fields by
\beq
\bold Q \equiv \Delta_G \cup \left(\cup_{a =1}^A\cR_a\right)  \ ,
\eeq
and the size of this charge set by $N\equiv |\bold Q|$. The index $i$ is used
to label the vector and the matter charges altogether.
The charges are thus labelled as $Q_{i=1, \cdots, N}$
and the associated hyperplanes, as $H_{i=1, \cdots, N}$.
Finally, given a singularity at $u=u_*$, let us denote by
$\bold H_{u_*}$ the set of hyperplanes on which $u_*$ lies,
\beq
\bold H_{u_*} \equiv \{\,H_i \,|\, u_* \in H_i \, \} \ ,
\eeq
and by $\bold Q_{u_*}$ the set of their associated charges,
\beq
\bold Q_{u_*} \equiv \{\, Q_i \,|\, u_* \in H_i \, \} \ .
\eeq

\subsubsection*{Genericity and Strong Regularity Criteria for $\eta$}

Before getting at the systematic approach to the
twisted partition function computation, let us first clarify the genericity
criterion we demand for the choice of $\eta$ in the formula~\eqref{loc}. It
is known that the final outcome of the JK-Res operation,
computed as the sum of JK residues over various singularities
of two types,~\eqref{JKsum1} and~\eqref{JKsum2},
does not depend on the choice of this fictitious vector $\eta$,
given that a genericity criterion is obeyed. This
demands that $\eta$ cannot be spanned by less than
$r$ charge vectors. It should be noted that each JK residue at a
given singularity may jump individually as $\eta$ crosses a wall, while
the sum~\eqref{loc} remains intact.

When we resort to the notion of flags, as is necessary
when degenerate singularities are present, it
turns out that such a genericity condition is not
quite enough. One finds that, when $\eta$ points along a
particular integer sum of charges, the flag prescription
gives too many residues. This can be seen clearly
in rank two nondegenerate examples such as $SU(3)$ pure Yang-Mills.
To avoid this additional subtlety, we require the Strong Regularity condition,
\beq\label{eta-gen}
\eta \notin {\rm Cone}_{\rm sing} \big\{\,{\Sigma}_{i \in \pi}\, Q_i \,|\,
\pi \subset \{1, \cdots, N\} \,  \big\} \ ,
\eeq
where $\rm Cone_{sing}$ of a set of vectors denotes the union of
the cones generated by any $r-1$ vectors in the set.
The meaning of the criterion~\eqref{eta-gen} will become clear
in the context of flags below.

\subsubsection*{JK-Positive Collection of the Charges}

We start by collecting all the subsets of $r$ independent charges,
\beq\label{jkpc}
\bold Q_{i_1, \cdots, i_r}=\{ Q_{i_1}, \cdots, Q_{i_r} \} \subset \bold Q \ ,
\eeq
such that ${\rm rk} (Q_{i_1} \cdots Q_{i_r}) = r$.
We further impose that the following positivity condition is obeyed,
\beq
\eta \in {\rm Cone} (Q_{i_1}, \cdots, Q_{i_r})
 \equiv \{\, \sum_{p=1}^r a_p Q_{i_p} \,|\, a_p \geq 0\,\} \ ,
\eeq
with respect to the given choice of $\eta$
that obeys the Strong Regularity criterion.
The classification of such ``JK-positive'' charge collections
is the starting point for the computation of the twisted partition function.

\subsubsection*{Singularities, Flags and JK Residues}
Since each JK-positive collection, $\bold Q_{i_1, \cdots, i_r}$,
is linearly independent, the corresponding hyperplane collection,
\beq
\bold H_{i_1, \cdots, i_r} = \{H_{i_1}, \cdots, H_{i_r}\} \ ,
\eeq
leads to codimension-$r$ singularities, $u_* \in \bigcap\limits_{p=1}^r H_{i_p}$.
In general, the $r$ hyperplanes give rise to more than one
singularities since the hyperplane equations,~\eqref{hv}
and~\eqref{hm}, only need to hold modulo $2 \pi i $. Then, given the
periodicity of the $u$ variables,
\beq
u_k\sim u_k+2\pi i \ ,
\eeq
there arise finitely many discrete singularities for each JK-positive collection.

For each of the singularities $u_*$ obtained
from the collection $\bold Q_{i_1, \cdots, i_r}$,
we must proceed to construct the corresponding flags that
satisfy themselves yet another positivity criterion.
Note first that the singularity $u_*$ can be degenerate
so that the charge set $\bold Q_{u_*}$ may contain more charges
than the $r$ charges $Q_{i_1}, \cdots, Q_{i_r}$.
Once $\bold Q_{u_*}$ is found, we may forget about
$\bold Q_{i_1, \cdots, i_r}$ completely and consider
the set $\mathcal{FL}(\bold Q_{u_*})$ of flags
\beq
\cF=\left[\cF_0 =\{0\} \subset \cF_1 \subset \cdots \subset \cF_r =\IC^r \right]
 \ , \quad {\rm dim}\,\cF_k = k \ ,
\eeq
such that the vector space $\cF_k$ at each level $k$ is
spanned by $\{Q_{j_1}, \cdots, Q_{j_k}\}$, with $k =1, \cdots, r$,
where the ordered set $\bold B(\cF) \equiv \{Q_{j_1}, \cdots, Q_{j_r} \}$
is a subset of $\bold Q_{u_*}$.
As one last piece of ingredient,
we construct the vector $\kappa_k^\cF$ at each level $k$ as,
\beq
\kappa_k^\cF = \sum_{Q_i \in \bold Q_{u_*} \cap \cF_k} Q_i \ ,
\eeq
and define the sign factor,
\beq\label{sf}
\nu(\cF)={\rm sign}\, {\rm det}\, (\kappa_1^\cF \cdots \kappa_r^\cF) \ .
\eeq

The JK residue at $u_*$ is then given as,
\beq
\underset{u=u_*}{\text{ JK-Res}}\,_{\eta:\bold Q_{u_*}}
= \sum_{\cF\in \mathcal{FL}^+(\bold Q_{u_*}, \eta)} \nu(\cF)
\, \underset{\cF} {\rm Res} \ ,
\eeq
where $\mathcal{FL}^+(\bold Q_{u_*}, \eta) = \{\, \cF \in
\mathcal{FL}(\bold Q_{u_*}) \,|\, \eta \in
{\rm Cone}(\kappa_1^\cF, \cdots, \kappa_r^\cF) \,\}$,
and the {\it iterated residue} associated to $\cF$,
\beq
\underset{\cF}{\rm Res} \ ,
\eeq
is defined in terms of the basis $\bold B(\cF)$ as follows.
Given an $r$-form $\omega=\omega_{1\cdots r} \,{\rm d}
u_1\wedge \cdots \wedge {\rm d} u_r$, the following coordinate change
is made,
\beq
\tilde u_k = Q_{j_k} \cdot u \, \quad k=1, \cdots, r \ ,
\eeq
with respect to the basis $\bold B(\cF)$,
and the $r$-form $\omega$ is rewritten in terms of
the new coordinates,
\beq
\omega = \tilde \omega_{1\cdots r}\,
{\rm d}\tilde u_1 \wedge \cdots \wedge {\rm d}\tilde u_r \ .
\eeq
Then the iterated residue is defined as,
\beq\label{ir}
\underset{\cF}{\rm Res}\; \omega = \underset{\tilde u_r
=\tilde u_{*r}}{\rm Res} \cdots \underset{\tilde u_1
=\tilde u_{*1}}{\rm Res} \, \tilde \omega_{1\cdots r} \ ,
\eeq
where the $\rm Res$ at each level $k$ on the RHS is
the usual single-variable residue operation, performed by
regarding any other variables as a generic constant.

\subsubsection*{Twisted Partition Function}
Once the singularities $u_*$ of the integrand are classified,
together with the associated flags $\mathcal{FL}^+(\bold Q_{u_*}, \eta)$
with respect to a given $\eta$, the twisted partition function
is obtained in a straightforward manner. All that is left
is to compute the iterated residues following eq.~\eqref{ir},
one for each pair of a singularity and a flag, and gather those
residues together, taking into account the sign factor~\eqref{sf}.

\subsection{Cancelation of Asymptotic Residue for Pure Yang-Mills}\label{B}

Let us show that for pure Yang-Mills theories, the localization
leads to no residue contributions from the asymptotic boundary of $(\IC^*)^r$.
Pure Yang-Mills quantum mechanics with a simple gauge group, does
not have a Fayet-Iliopoulos constant that can control such
an asymptotic residue via $Q_\infty=-\zeta$. Thus, we must go back to
the basics of localization to figure out whether the
asymptotic poles contribute or not.

Tracing back to localization derivation of Ref.~\cite{HKY},
we find that $D$-integral is performed a little differently
at internal poles and at asymptotic poles. When $\zeta$
is present and can be scaled up with $e^2\zeta$ finite, $D$-integral
gives a step function with the argument proportional to $\zeta$ in
such a way that, depending on JK positivity test with a given $\eta$,
it may or may not contribute. Naively keeping $\zeta$ finite or
scaling it down to zero, this step function becomes an error function,
so that, with $\zeta=0$ effectively, the $D$-integral gives $1/2$
relative to when it actually contributed by passing the JK test.
Now in our case, in the absence of $\zeta$ to begin with, the end result
is the same by taking the $D$-integral from field theory at face value
and taking the principal value along $D=0$ pole in the integrand.

Potential contributions thus have the form,
\begin{equation}
\frac{1}{4\pi i} \prod_{p=1}^{r-1} \Theta(-Q_{i_p}\cdot\delta_p)\times
\int_{\partial \IC^*} {\rm d}\mathfrak u\; \mathfrak{g}_{Q_{i_1}, Q_{i_2},\dots, Q_{i_{r-1}}}(\mathfrak u;z)
\end{equation}
where we have chosen some  shift $\delta_p$ of $D$-contours for the first
$r-1$ integral, resulting in the step functions, while
the last integral on the remaining variable, denoted as $\mathfrak u$, is
over the circles around the two asymptotic boundaries
of the cylinder $\IC^*$ at Re$\,\mathfrak u=\pm \infty$. The integrand $\mathfrak{g}$
for the $\mathfrak u$ integration is obtained from the original integrand
$g$ upon performing $r-1$ iterated residue computations.
What we wish to show is that, for pure $\CN=4,8,16$ Yang-Mills
quantum mechanics, this remaining integral on the right vanishes
on its own, in such a way that the complicated step functions in
front becomes irrelevant.

A prototype is $\CN=4$ pure $SU(2)$ theory, for which this
expression simplifies to
\begin{equation}\sim \frac{1}{4\pi i} \int_{\partial \IC^*} {\rm d}\mathfrak u
\;{ \mathfrak g}(\mathfrak u;z) =
\frac{1}{4\pi i} \int_{\partial \IC^*} {\rm d}\mathfrak u
\;\frac{\sinh^2(\mathfrak u)}{\sinh(\mathfrak u+z/2)\sinh(\mathfrak u-z/2)} \ ,
\end{equation}
where $\mathfrak u$ is the same as $u$ as no iterated residue computation
is performed in this case.
The integral consists of the two boundaries $\mathfrak u=\pm\infty+i\phi$
but with mutually opposite orientations for $d\phi$. On the other
hand, the integrand approaches the finite value of 1, independent
of $z$, at both of these two boundaries, giving us
\begin{equation}
\frac{1}{4\pi i}\left(2\pi i-2\pi i\right)=0
\end{equation}
It should be now clear what happens for a general gauge group,
possibly with additional adjoint chirals.

For this, let us start with $\CN=4$ and consider a co-dimension
$r-1$ singularity $l_*$ with the associated charge set
${\bold Q}_{l_*}$. Generally, the $\{Q_{i_p}\}$ above, basis for
a flag, is a subset of ${\bold Q}_{l_*}$. For a given $l_*$,
there could be multiple contributing iterated residues but
since the vanishing argument does not depend on such details,
let us consider one arbitrarily choice. When the $l_*$ is
nondegenerate, of course we have ${\bold Q}_{l_*}=\{Q_{i_p}\}$.

Let us denote by $\Delta_{l_*}^{\rm const}$ the set of root vectors~$\alpha$
for which $\alpha \cdot u$ is a constant over the entire line $l_*$.
Then, $\Delta_{l_*}^{\rm const}$ includes ${\bold Q}_{l_*}$ as a subset, responsible
for the singular line $l_*$. 
We group the one-loop determinant $g(u)$ into two factors as
\bea
g_{l_*}(u)\equiv \prod_{\alpha\in\Delta_{l_*}^{\rm const}}
\frac{\sinh^2(\alpha\cdot u/2)}{\sinh((\alpha\cdot u-z)/2)\sinh((\alpha\cdot u+z)/2)} \ , \eea
and
\bea f_{l_*}(u)\equiv \prod_{\alpha\in\Delta\setminus \Delta_{l_*}^{\rm const}}
\frac{\sinh^2(\alpha\cdot u/2)}{\sinh((\alpha\cdot u-z)/2)\sinh((\alpha\cdot u+z)/2)} \ , \eea
so that
\bea g(u)=g_{l_*}(u) \times f_{l_*}(u) \ . \eea
Note that the charges involved in $f_{l_*}(u)$ are such that $\alpha \cdot u$ is
not fixed to a number at $l_*$, leading to a free direction in $u$-space;
We denote this residual direction by a vector $s_{\rm res}$ and parameterize
the singular line $l_*$ by the holomorphic variable $\mathfrak u$ as
\beq
u=s_{\rm res}\, \mathfrak u + u_* \ , \qquad \mathfrak u \in \IC \ ,
\eeq
where $u_*$ is a point in $l_*$, chosen arbitrarily.

If only $r-1$ hyperplanes collide along $l_*$,
all the poles involved in the first $r-1$ integrations are a simple pole
and $f_{l_*}(u)$ is merely evaluated at $l_*$.
Thus, when it comes to the last integration,
$f_{l_*}(u)$ becomes a function only of $\mathfrak u$,
\bea
f_{l_*}(u) \quad\xrightarrow[]{{\text{restricted~to}}~l_*}\quad
\tilde f_{l_*}(\mathfrak u):=f_{l_*} (s_{\rm res}\, \mathfrak u + u_*) \ ,
\eea
along the singular line $l_*$.
Note in particular that
\bea
\tilde f_{l_*}(\mathfrak u\rightarrow+\infty)=1=\tilde f_{l_*}(\mathfrak u\rightarrow-\infty) \ ,
\eea
at the asymptotic infinity.
The final integrand $\mathfrak g(\mathfrak u)$ is then the product of
$\tilde f_{l_*}(\mathfrak u)$ and the value of the residue from $g_{l_*}$,
say, $\tilde g_{l_*}$.
Since the latter is independent of $\mathfrak u$, we thus have
\bea
\mathfrak g (\mathfrak u\rightarrow+\infty)=\tilde g_{l_*}=\mathfrak g(\mathfrak u\rightarrow-\infty) \ ,
\eea
so that the residues at the two asymptotic poles sum to zero.

When more than $r-1$ charges collide at $l_*$, say $r-1+n$ of them,
only a slight modification occurs. Because of the presence of degenerate
poles, residue operation could take further contributions
from derivatives of $g_{l_*}(u)$ and $f_{l_*}(u)$, so the final integrand
$\mathfrak g(\mathfrak u)$ takes the form,
\bea
\sum_{k=0}^n \sum\limits_{a=1}^{a_k} \tilde g^{(n-k);a}_{l_*}\times \tilde f^{(k);a}_{l_*} (\mathfrak u) \ ,
\eea
where $\tilde g^{(n-k);a}_{l_*}$ and $\tilde f^{(k);a}_{l_*} (\mathfrak u)$ are appropriate results of the first $r-1$
residue operations. The latter is in particular obtained after
taking $k$ number of derivatives on $f_{l_*}(u)$, where the superscript,
$a=1,\cdots, a_k$, labels numerous different terms with
the same total number of derivatives on $f_{l_*}(u)$.
Again, only $\tilde f^{(k);a}_{l_*} (\mathfrak u)$'s carry $\mathfrak u$-dependence.
Since $f_{l_*}(u)$ is
asymptotically constant up to exponentially small corrections
$\sim e^{-|c \,\cdot\,{\rm Re}\,\mathfrak u|}$, their derivatives
vanish asymptotically and the only surviving
piece as ${\rm Re}\,\mathfrak u \rightarrow \pm\infty$ is the $k=0$ piece,
\bea
\mathfrak g(\mathfrak u\rightarrow\pm\infty)
=\tilde g^{(n)}_{l_*} \times \tilde f_{l_*} (\mathfrak u \rightarrow \pm\infty)= \tilde g^{(n)}_{l_*}\ ,
\eea
which again cancel against each other.

Generalization to larger supersymmetry is straightforward. The
underlying reason behind this vanishing argument is that,
for a given flavor and $R$-charge, a pair of mutually
opposite gauge charges contribute together to the original integrand $g(u)$.
When only adjoint $\cN=4$
chiral multiplets are present, we can split the integrand as
a product of
\bea
\frac{\sinh((\alpha\cdot u+q z+F\cdot \mu)/2)
\sinh((-\alpha\cdot u+q z+F\cdot \mu)/2)}{\sinh((\alpha\cdot u+(q-1)z
+F\cdot \mu)/2)\sinh((-\alpha\cdot u+(q-1)z+F\cdot \mu)/2)} \ ,
\eea
for some $q$'s and $F$'s. This approaches to 1 at the asymptotic
regions of $(\IC^*)^r$, regardless of $q$ and $F$, which lets
us proceed in the same manner as in $\cN=4$ case above. Because
$\cN=8$  and $\cN=16$ can be constructed by adding either one or three
sets of adjoint chirals and a constraining superpotential appropriately,
the argument for the cancelation of asymptotic residues still holds
for these cases.

\section{Elliptic Weyl Elements}\label{C}

\subsection{Classical Groups}

An elliptic element $w$ of Weyl group $W$ is defined
by absence of eigenvalue 1 in the canonical $r$-dimensional
representation of $W$ on the weight lattice.
For $SU(N)$, the Weyl group is the permutation group $S_N$ and
the relevant representation for it is the $(N-1)$-dimensional irreducible
one. Of elements of $S_N$, the only elliptic Weyl are the
fully cyclic ones, say, $(123\cdots N)$ and the permutations
thereof, where $(\dots)$  represents the cyclic permutation.

For $SO(2N)$, $SO(2N+1)$, and $Sp(N)$ groups, the Weyl
groups are $S_N$ semi-direct-product with $(Z_2)^{N-1}$,
$(Z_2)^{N}$, and $(Z_2)^{N}$, respectively. The elements
can be therefore represented as follows
$$
w=(ab\dot c \dot d \dots)(klm\dot n \dots)\cdots
$$
where $(\dots)$ again represents a cyclic permutation and dots above a number
indicate a sign flip. For example $(12\dot 3)$ represents the element,
$$
\left(\begin{array}{ccc} 1 &0&0\\ 0&1&0\\ 0&0&-1\end{array}\right)\cdot
\left(\begin{array}{ccc} 0 &1&0\\ 0&0&1\\ 1&0&0\end{array}\right) \ .
$$
In this form, the above $(Z_2)^{N-1}$ for $SO(2N)$ means
that the total number of sign flip has to be even.
Since the determinant factorizes upon the above decomposition
of $w$, this should be true for each cyclic component.
It is fairly easy to see that this requires  each
cyclic component of $w$ to have an odd number of sign
flips.

Let us list the relevant conjugacy classes of
classical groups, by listing typical elements in each class.
We list up to rank five, except for $SU(N)$,
\begin{itemize}
\item $SU(N)$
$$(123\cdots N)$$
\item $SO(4)$
$$(\dot 1)(\dot 2)$$
\item $SO(5)$ and $Sp(2)$
$$(1\dot 2),\quad(\dot 1)(\dot 2)$$
\item $SO(6)$
$$(1\dot 2)(\dot 3)$$
\item $SO(7)$ and $Sp(3)$
$$(\dot 1\dot 2\dot 3),\quad (12\dot 3),\quad (1\dot 2)(\dot 3),
\quad (\dot 1)(\dot2)(\dot 3)$$
\item $SO(8)$
$$(\dot 1\dot 2\dot 3)(\dot 4),\quad(12\dot 3)(\dot 4),\quad (1\dot 2)(3\dot 4),
\quad (\dot 1)(\dot 2)(\dot 3)(\dot 4)$$
\item $SO(9)$ and $Sp(4)$
$$(1\dot 2\dot 3\dot 4),  \quad (1 2 3\dot 4),\quad (\dot 1\dot 2\dot 3)(\dot 4),
\quad(12\dot 3)(\dot 4),\quad (1\dot 2)(3\dot 4),\quad
(1\dot 2)(\dot 3)(\dot 4), \quad (\dot 1)(\dot 2)(\dot 3)(\dot 4)$$
\item $SO(11)$ and $Sp(5)$
$$(\dot 1\dot 2\dot 3\dot 4 \dot 5),  \quad (1 2 \dot 3\dot 4\dot 5),
\quad  (1 2 3 4\dot 5), \quad (1\dot 2\dot 3\dot 4)(\dot 5),\quad (1  2 3\dot 4)(\dot 5),\quad
(\dot 1 \dot 2\dot 3)(4\dot 5), \quad( 1  2\dot 3)(4\dot 5),$$
$$(\dot 1\dot 2 \dot 3)(\dot 4)(\dot 5),\quad  ( 1 2 \dot 3)(\dot 4)(\dot 5),
\quad (1\dot 2)(3\dot 4)(\dot 5),\quad (1\dot 2)(\dot 3)(\dot 4)(\dot 5),
\quad (\dot 1)(\dot 2)(\dot 3)(\dot 4)(\dot 5)$$

\end{itemize}

\subsection{Exceptional Groups: $G_2$ and $F_4$}

For exceptional gauge groups, the Weyl symmetries are more involved.
For $G_2$, the dihedral group $D_6$ with 12 elements is the Weyl group.
This group is a symmetry group of $6$-gon with 6 rotations and 6
reflections. Reflections cannot be elliptic, as it always leaves
a direction intact. Of remaining rotations, $Z_6$, five proper rotations
are all elliptic Weyl elements.

For $F_4$, the Weyl group is semidirect product of $S_4$ which permutes
the four basis, $e_{1,2,3,4}$, $(Z_2)^3$ which flips an even number of $e_i$'s,
and $S_3$
generated by
\begin{eqnarray}
\alpha\equiv \frac{1}{2}\left(\begin{array}{rrrr}
1  &1  &1  &1\\
1  &1  &-1 &-1\\
1  &-1 &1  &-1\\
1  &-1 &-1 &1\end{array}
\right) \ ,
\qquad
\beta\equiv\left(\begin{array}{rrrr}
-1  &0  &0  &0\\
0  &1  &0 &0\\
0  &0 &1  &0\\
0  &0 &0 &1\end{array}
\right) \ .
\end{eqnarray}
Alternatively, this can be thought of as a semidirect product of
$(Z_2)^4$ that flips signs of each of $e_i$, the permutation $S_4$,
and the $Z_3$ consisting of the identity and the two matrices,
\begin{eqnarray}
\gamma\equiv \alpha\beta = \frac{1}{2}\left(\begin{array}{rrrr}
-1  &1  &1  &1\\
-1  &1  &-1 &-1\\
-1  &-1 &1  &-1\\
-1  &-1 &-1 &1\end{array}
\right) ,~~
\gamma^2 = \gamma^{-1}=\frac{1}{2}\left(\begin{array}{rrrr}
-1  &-1  &-1  &-1\\
1  &1  &-1 &-1\\
1  &-1 &1  &-1\\
1  &-1 &-1 &1\end{array}
\right) .
\end{eqnarray}
In other words, the Weyl group of $F_4$ is a semi-direct
product of $Z_3$, generated by $\gamma$, with the Weyl group
of $SO(9)$, with the cardinality $3\cdot 2^4\cdot 4! =1152$.
Since the determinant of $\gamma$ is unit, the expression
we are after for $\Omega_{\cN=4}^{F_4}({\bf y})$ is
$$
\frac{1}{1152}\left(\sum'_{w}\frac{1}{{\rm Det}\left(y^{-1}-y\cdot w\right)}
+\sum''_{w}\frac{1}{{\rm Det}\left(y^{-1}\gamma-y\cdot w\right)}
+\sum'''_{w}\frac{1}{{\rm Det}\left(y^{-1}\gamma^{-1}-y\cdot w\right)}\right) \ ,
$$
where the three sums are over subsets of $SO(9)$ Weyl group,
restricted to those $w$'s such that ${\rm Det}(\gamma^n-w)\neq 0$
for $n=0,\pm 1$, respectively.

\end{document}